\documentclass[sigconf,screen, authorversion]{acmart}

\usepackage{algorithm}
\usepackage[commentColor=blue]{algpseudocodex}
\usepackage{graphicx,subcaption}
\usepackage{color}
\AtBeginDocument{%
  \providecommand\BibTeX{{%
    \normalfont B\kern-0.5em{\scshape i\kern-0.25em b}\kern-0.8em\TeX}}}

\copyrightyear{2025}
\acmYear{2025}
\setcopyright{acmlicensed}
\acmConference[PPoPP '25]{The 30th ACM SIGPLAN Annual Symposium on Principles and Practice of Parallel Programming}{March 1--5, 2025}{Las Vegas, NV, USA}
\acmBooktitle{The 30th ACM SIGPLAN Annual Symposium on Principles and Practice of Parallel Programming (PPoPP '25), March 1--5, 2025, Las Vegas, NV, USA}
\acmDOI{10.1145/3710848.3710890}
\acmISBN{979-8-4007-1443-6/25/03}

\newcommand{\punt}[1]{}
\newcommand{\cmnt}[1]{}

\definecolor{xxxcolor}{rgb}{0.8,0,0}
\newcommand{\func}[1]{\texttt{#1}}

\newtheorem{theorem}{Theorem}

\newtheorem{property}[theorem]{Property}

\newcounter{history}

\newtheorem{assumption}[theorem]{Assumption}
\newcommand{\secref}[1]{Section~\ref{sec:#1}}
\newcommand{\figref}[1]{Figure~\ref{fig:#1}}

\newcommand{\asmref}[1]{Assumption~\ref{asm:#1}}

\newcommand{\lineref}[1]{line~\ref{lin:#1}}
\newcommand{\algoref}[1]{Algorithm~\ref{algo:#1}}

\newcommand{\ignore}[1]{}
\newcommand{\myparagraph}[1]{\noindent\textbf{#1}}

\graphicspath{ {figs/} }

\begin{document}

%%
%% The "title" command has an optional parameter,
%% allowing the author to define a "short title" to be used in page headers.
\title{Publish on Ping: A Better Way to Publish Reservations in Memory Reclamation for Concurrent Data Structures}
%%
%% The "author" command and its associated commands are used to define
%% the authors and their affiliations.
%% Of note is the shared affiliation of the first two authors, and the
%% "authornote" and "authornotemark" commands
%% used to denote shared contribution to the research.
% \author{Ben Trovato}
% \authornote{Both authors contributed equally to this research.}
% \email{trovato@corporation.com}
% \orcid{1234-5678-9012}
% \author{G.K.M. Tobin}
% \authornotemark[1]
% \email{webmaster@marysville-ohio.com}
% \affiliation{%
%   \institution{Institute for Clarity in Documentation}
%   \streetaddress{P.O. Box 1212}
%   \city{Dublin}
%   \state{Ohio}
%   \country{USA}
%   \postcode{43017-6221}
% }

\author{Ajay Singh}
\affiliation{%
  \institution{University of Waterloo}
  \streetaddress{}
  \city{}
  \country{Canada}}
  \orcid{0000-0001-6534-8137}
\email{ajay.singh1@uwaterloo.ca}

\author{Trevor Brown}
\affiliation{%
  \institution{University of Waterloo}
  \city{}
  \country{Canada}}
  \orcid{0000-0002-0074-1031}
\email{trevor.brown@uwaterloo.ca}

%%
%% By default, the full list of authors will be used in the page
%% headers. Often, this list is too long, and will overlap
%% other information printed in the page headers. This command allows
%% the author to define a more concise list
%% of authors' names for this purpose.
% \renewcommand{\shortauthors}{Trovato and Tobin, et al.}

%%
%% The abstract is a short summary of the work to be presented in the
%% article.
\begin{abstract}

Safe memory reclamation techniques that utilize per read reservations, such as hazard pointers and hazard eras, often cause significant overhead in traversals of linked concurrent data structures. This is primarily due to the need to announce a reservation, and fence to make it globally visible (and enforce appropriate ordering), before each read. In real world read-intensive workloads, this overhead is amplified because, even if relatively little memory reclamation actually occurs, the full overhead of reserving records before use is still incurred while traversing data structures.

In this paper, we propose a novel memory reclamation technique by combining POSIX signals and delayed reclamation, introducing a publish-on-ping approach. 
This method eliminates the need to make reservations globally visible before use. Instead, threads privately track which records they are accessing, and share this information on demand with threads that intend to reclaim memory. 
The approach can serve as a drop-in replacement for hazard pointers and hazard eras. Furthermore, the capability to retain reservations during traversals in data structure operations and publish them on demand facilitates the construction of a variant of hazard pointers (EpochPOP). This variant uses epochs to approach the performance of epoch-based reclamation in the common case where threads are not frequently delayed (while retaining the robustness of hazard pointers).

Our publish-on-ping implementations based on hazard pointers and hazard eras, when applied to various data structures, exhibit significant performance improvements. The improvements across various workloads and data structures range from 1.2X to 4X over the original HP, up to 20\% compared to a heavily optimized HP implementation similar to the one in the Folly open-source library, and up to 3X faster than hazard eras. 
EpochPOP delivers performance similar to epoch-based reclamation while providing stronger guarantees.

\end{abstract}

%%
%% The code below is generated by the tool at http://dl.acm.org/ccs.cfm.
%% Please copy and paste the code instead of the example below.
%%


% \ccsdesc[500]{Do Not Use This Code~Generate the Correct Terms for Your Paper}
% \ccsdesc[300]{Do Not Use This Code~Generate the Correct Terms for Your Paper}
% \ccsdesc{Do Not Use This Code~Generate the Correct Terms for Your Paper}
% \ccsdesc[100]{Do Not Use This Code~Generate the Correct Terms for Your Paper}

%%
%% Keywords. The author(s) should pick words that accurately describe
%% the work being presented. Separate the keywords with commas.

% \received{20 February 2007}
% \received[revised]{12 March 2009}
% \received[accepted]{5 June 2009}

%%
%% This command processes the author and affiliation and title
%% information and builds the first part of the formatted document.
\maketitle

\section{Introduction}
\label{introduction}
% Be crisp, to the point and natural.
% \textcolor{blue}{How should I answer the question that asymmetric memory fencing overcomes the overhead?}

% \subsection{The Problem}
There is a rich literature of concurrent data structures that can be used as building blocks for concurrent software.
However, many concurrent data structure designs assume automatic garbage collection, and cannot be used (without modification) in % to be practically useful in
environments with no garbage collection, such as C/C++.
In such environments, concurrent data structures typically must be paired with safe memory reclamation (SMR) algorithms~\cite{brown2015reclaiming,michael2004hazard,detlefs2001lock,gidenstam2008efficient,hart2007performance,cohen2015automatic,balmau2016fast,alistarh2018threadscan,cohen2018every,wen2018interval,dice2016fast,herlihy2005nonblocking,blelloch2020concurrent,ramalhete2017brief,cohen2015efficient, alistarh2014stacktrack, dragojevic2011power, alistarh2017forkscan, braginsky2013drop, nikolaev2019hyaline, kang2020marriage, nikolaev2020universal, nikolaev2021brief, nikolaev2019hyaline, singh2021nbr, singh2023efficient, singhTPDS2023NBRP, jung2023applying} to prevent %provide safety from
use-after-free errors.
As an example of a use-after-free error, consider a traversal of a linked-list based set that does not use locks or other synchronization in its traversals: While one thread deletes and subsequently reclaims a node another thread could concurrently access it, potentially leading to a segmentation fault. % in programs.

%These reclamation algorithms are commonly referred to as \textit{safe memory reclamation} algorithms~\cite{all}.
Many of the most common SMR algorithms are \textit{pointer based}.
The most notable example, the celebrated hazard pointer (HP) algorithm~\cite{michael2004hazard}, has received a Dijkstra award, and is set to be included in the C++26 standard~\cite{michael2017hazard}.
Unfortunately, in HPs, %this algorithm to avoid use-after-free errors due to the concurrent reclamation, 
whenever a thread encounters a new shared object, such as a node in a list, the thread must \textit{reserve} a pointer to the node by (1) storing it in a single-writer multi-reader (SWMR) slot in a shared array, (2) executing memory fence instruction(s) to \textit{publish} this reservation, making it visible to other threads (and preventing instruction reordering), and (3) re-reading a pointer from which this node was encountered to verify that the reserved node is still reachable at some time after the reservation was published.

The need to fence every time a new node is encountered can cause high overhead in linked data structures.
This overhead is even more pronounced in common read-intensive workloads wherein, even though memory reclamation may be infrequent, and relatively lightweight, the overhead that fencing imposes on read-only operations is not. % due to the reclamation related step is incurred during the traversals.
A more recent technique, called hazard eras (HEs)~\cite{ramalhete2017brief}, uses monotonically increasing global timestamps and reservation of timestamps, instead of pointers. %, to protect reads. %at the time of reads.
Roughly speaking, HEs is more coarse grained, as a timestamp represents many nodes. 
The use of timestamps reduces how often memory fences are needed.
For example, if a thread is about to reserve a timestamp that it has already reserved (because a new node being accessed corresponds to the same timestamp as a previously accessed node), the previously published timestamp can be reused (without additional fencing).
%A fence is required only if the timestamp has changed since it was last reserved and published. It reduced the memory fences and can be used as a drop-in replacement for hazard pointers, but the overhead still depended on how frequently the global timestamp is set to change.
However, the overhead of HEs can still be substantial. 
%Various other prior techniques provided operating system and hardware dependent ways to mitigate this overhead, but they are not necessarily compatible with the same set of data structures as HPs and HEs, or with all systems. %, which may not be easy to integrate into we believe are not as easy to integrate. 

Another alternative is to use fast epoch based reclamation (EBR) algorithm~\cite{hart2007performance,fraser2004practical,mckenney1998read}.
However, these techniques are not robust: A delayed thread can prevent all threads from reclaiming memory (forever) resulting in unbounded memory consumption. % of concurrent data structures.
Subsequent reclamation techniques like NBR~\cite{singh2021nbr, singhTPDS2023NBRP}, VBR~\cite{sheffi2021vbr} and IBR~\cite{wen2018interval} provide a variety of robustness guarantees, but they have other trade-offs in, for example, ease of integration, applicability to classes of data structures, assumptions regarding whether memory pages are returned to the operating system, and so on.
NBR requires data structure operations to have a particular structure, consisting of read- and write-phases, with specific requirements in each phase. %, and relies on POSIX signals to achieve robustness. %and use checkpoints per operation, where a neutralized thread during traversals could jump back to in response to signals from reclaimers.
Optimistic techniques like VBR %and other reclaimers in this category
require a type preserving allocator, and cannot allow memory pages to be returned to the operating system (without tricks involving trapping and ignoring segmentation faults).
Moreno and Rocha designed a custom allocator that leverages the virtual memory subsystem to solve this problem by customizing page fault handling~\cite{moreno2023releasing}.
However, these techniques cannot be used as %Many of the subsequent techniques might not be easily used as a 
drop-in replacements for hazard pointers without making sacrifices in ease of use, portability, or applicability to certain data structures. %, either in the form of operating system and hardware support, supported. 

% \subsection{The Contribution}
In this paper, we propose a technique called \textit{publish-on-ping (POP)} which can be used to design fast and robust drop-in replacements for pointer based techniques, such as hazard pointers and hazard eras.
The POP technique can be applied to hazard pointers to significantly improve performance, resulting in an algorithm we call \textit{HazardPtrPOP}.
Perhaps surprisingly, EBR can also be incorporated into HazardPtrPOP, allowing different threads to \textit{simultaneously} operate in epoch-based and hazard pointer modes. %\textcolor{red}{where both operate in a synchronization-less dual mode},
This results in an algorithm called \textit{EpochPOP} that achieves performance similar to EBR, while retaining the robustness guarantees of HazardPtrPOP.
%Perhaps surprisingly, this technique can be used in conjunction with epoch based reclamation to also help solve the speed vs robustness choice programmers have to make without sacrificing ease in programming or requiring type preservation or architectural dependence.

The key to safety from use-after-free errors in most pointer based techniques is that threads traversing a data structure eagerly reserve and publish (making the reservations visible to all threads) before accessing nodes during traversals, and threads reclaiming objects must first scan all the reservation to avoid freeing reserved objects.
It is highly pessimistic to publish reservations (and fence) %requires executing a memory fence at every read
just because there \textit{might be} some concurrent reclaimer thread about to free some nodes.
Even if there is very little (or no) reclamation in a given workload, traversals must pay this cost for every node visited.
In this paper, we ask the question: What if each traversal could publish its set of reservations \textit{precisely} when a reclaimer is about to scan threads' reservations? %free are about to reclaim. We call this paradigm of publishing reservations on demand as publish-on-ping (pop).

As in the recent NBR algorithm, POSIX signals are the key to our POP technique. % is component of our approach are the POSIX signals. 
However, signals elicit different behaviour from the recipient in POP than in NBR.
In NBR, %threads accumulate garbage in local buffers, and when a buffer becomes full,
a thread that wants to reclaim memory sends signals to all other threads, causing those threads to discard all pointers they hold and \textit{restart} their current traversals (unless they have already performed writes to the data structure).
%This can necessitate substantial changes to the control flow of an existing data structure.
In POP, a thread that wants to reclaim memory sends signals to all other threads (as in NBR), but a thread receiving a signal does not need to restart its operation or otherwise change its control flow.
Rather, it simply publishes its reservations (which it had been tracking \textit{locally} until this point), increments a SWMR counter that tracks how many times it has published reservations, and issues a single memory fence.
Once all threads have published their reservations, the reclaimer can scan them and free nodes that are not reserved.
In a nutshell, this is how HazardPtrPOP works. %is implemented.
%the signal does not cause those threads to not have to change their control flow
%In contrast, POP does not require any changes to the control flow of a 
%is Threads instead of eagerly publishing their reservations keep them locally until a thread attempting to reclaim its retired nodes signals every thread to publish the reservations, upon which all threads execute a signal handler to publish their reservations, and the reclaimer then proceeds to scan and free nodes that are not reserved.

% So, for instance, when applied to hazard pointers this technique works as follows: during traversals all threads 1) save the nodes (reserve) locally without executing a memory fence, 2) verify that the reserved node was not deleted at the time it was reserved. These steps are repeated until it is established that the reserved node was not deleted concurrently. Before reclaiming, threads (wlog., say T1) signal (ping) other threads, upon receiving the ping, all threads respond by invoking a signal handler where they write their local reservations to SWMR slots and execute a memory fence to make them globally visible timely. When every thread completes execution of its signal handler, T1, similarly to hazard pointers, scans all the reservations and free objects that were not reserved.

EpochPOP builds on HazardPtrPOP as follows.
%We also use HazardPtrPOP to implement a robust epoch based reclamation algorithm.
In the common case, threads execute data structure operation largely as they would in epoch based reclamation~\cite{hart2007performance,fraser2004practical,mckenney1998read}.
A data structure operation begins by reading a global epoch number (a timestamp), and announcing the value it read in a per-thread SWMR slot in a shared array, then proceeds as usual.
Nodes unlinked from the data structure are stored in a per-thread list. %per-epoch limbo-bags.
Periodically, threads increment the global epoch, and scan other threads' announcements to identify the oldest announced epoch. 
% a data structure operation will scan other threads' announcements, and if all threads have announced the current global epoch, the global epoch is incremented.
% Whenever the global epoch is incremented, 
Threads can free objects unlinked from the data structure in epochs older than the oldest announced epoch.
% (as in traditional EBR~\cite{?? debra}.
However, unlike traditional EBR, all threads also keep executing the steps of the HazardPtrPOP algorithm, locally reserving nodes before accessing them. %, like in publish on ping.
If threads announce the epochs sufficiently often, they are able to reclaim memory regularly to keep the size of the lists with unlinked objects below a user-specified threshold, then the POP mechanism is not needed at all!
But, if thread delays hinder frequent announcements of epochs, preventing reclamation of the lists, then POP is activated to enable reclamation. % takes too long to advance because of thread delays, 
%During reclamation, threads reclaims nodes as they normally do in epoch based reclamation, except when they suspect that some thread might be stalled preventing reclamation, threads use pop to reclaim the nodes.

The dual mode of operation in EpochPOP differs significantly from the fast-path/slow-path approach used in Qsense~\cite{balmau2016fast}.
In Qsense, all threads globally switch (together) between a hazard pointer based mode, and an epoch based mode.
This involves additional synchronization and background threads, and necessitates careful decision making about \textit{when} all threads should switch between modes.
In the epoch based mode, if a single thread is delayed, all threads must switch to the hazard pointer mode.

% The dual mode of operation in EpochPOP differs from the fast-path/slow-path approach used in Qsense~\cite{balmau2016fast}. In Qsense, threads use Cadence (an optimized Hazard Pointer algorithm) as the slow-path, which involves background processes pinned to each core to produce context switches by descheduling main data structure threads at regular intervals. Quiescence-based reclamation is used as the fast-path. This approach is intrusive, requiring additional background processes per core, leading to context-switch overhead even when threads are not reclaiming. Additionally, threads in Qsense can only reclaim in either the fast-path or slow-path at a time, synchronizing through a shared flag.

In contrast, EpochPOP does not require any global mode switching (or requisite tuning).
Threads can effectively run in both modes simultaneously.
(One reclaimer may observe a thread delay, while another does not, and the latter can simply continue reclaiming with epochs while the former pings all threads.) %, so one reclaimer that observes a delay may ping other threads, while another reclaimer sees does not. %, and %allows different threads to effectively run in different modes.
%A slow thread will only cause a reclaimer to ping other threads if the reclaimer has already accumulated too much garbage.
%so a slow thread will not necessarily cause all threads to cause all 
QSense and POP both avoid fencing after individual reservations, but the techniques are quite different, and QSense requires additional background threads (one per core), whereas POP only needs signals.

Other signal-based techniques, such as NBR\cite{singh2021nbr}, DEBRA+~\cite{brown2015reclaiming}, HP-BRCU~\cite{kim2024expediting}, VBR~\cite{sheffi2021vbr} and PEBR~\cite{kang2020marriage} require programmers to either insert checkpoints and carefully identify safe regions for restarting or to add custom recovery code in data structure operations.
While this may be straightforward in some cases, it can be challenging or even infeasible in others~\cite{Cohen_thesis16, sheffi2021vbr}.  

Overall, POP is no more difficult to use than hazard pointers and is easier to implement than existing signal-based techniques that force thread restarts, like NBR. Similar to other signal-based approaches, POP algorithms avoid signal overhead when threads are not actively reclaiming memory. 
However, unlike these techniques, POP algorithms do not require thread restarts, making them well-suited for long-running read operations.

\subsubsection{Outline:}
The remaining paper is organized as follows. 
In \secref{background}, we recall Hazard Pointer and EBR to understand the algorithms presented in this paper. 
Then, in \secref{algorithms} we describe our Publish on Ping based algorithms-- HazardPtrPOP, HazardEraPOP and EpochPOP-- in detail followed by a discussion on correctness. In \secref{experimentalEvaluation}, we evaluate and compare the performance of HazardPtrPOP, HazardEraPOP and EpochPop with some previous techniques. 
Finally, in \secref{relatedwork} we discuss some technique similar in goals as ours and conclude in \secref{conclusion}.

% \section{Model}
% \label{sec:model}
% We consider the standard asynchronous shared memory model, where $N$ \textit{threads} access shared memory \textit{objects} using primitive memory access instructions\textendash \textit{read}, \textit{write}, and atomic \textit{read-modify-write}. The shared memory objects are part of an abstract data structure which threads can access using pointer traversals from an \textit{entry} point in the data structure.
% We assume that threads can communicate to each other via inter-process-interrupts (IPI) within a bounded time. 
% Although, signals in the current implementation uses locks, similar to other operating system services such as memory allocation, in principle, it is possible to implement lock-free signals. 

% The utilization of POSIX signals gives us access to the underlying scheduler, which we leverage to relax the traditional asynchronous memory model. In the traditional asynchronous memory model, learning the execution state of a thread (i.e., whether it has terminated or is merely delayed) is not possible.
% However, with POSIX signals, we can learn if a thread has terminated or is delayed arbitrarily.
% Specifically, when a thread sends an IPI, and the receiver has terminated, the sender learns the execution state of the target thread as an error message is returned to it. When a thread is delayed arbitrarily, like due to context switching, modern schedulers are reasonably dependable in ensuring that a signal handler will be executed within a bounded time.

\section{Background}
\label{sec:background}

A data structure object from its allocation to its freeing back to the underlying allocator can be in the following different states. 
\textit{reachable}: when the object after being allocated memory is inserted in to the data structure and threads could access it.
\textit{deleted}: when the object is logically marked for removal from the data structure.
% \textit{unreachable}: when the object is physically removed from the data structure such that no threads starting their traversal from its entry point (head in list o roots in trees) could reach the object.
\textit{retired}: when the object has been removed from the data structure, but yet not freed. Threads which gained access to the object before it was removed could still have access to it.
\textit{free}: when the object has been returned to the allocator and could be recycled for subsequent allocation requests.

\paragraph{Terminology}
A data structure object is said to be \textit{safe} for reclamation if it is in the retired state and no thread could access it. Otherwise, it is \textit{unsafe} for reclamation.
Typically, in \textit{deferred memory reclamation}, every thread has a local list, called \textit{retire list}, to which they add the \textit{retired} objects until they are safe to be freed.
This ensures retired objects are safe to be freed and amortizes reclamation overhead. 
When a retire list reaches a threshold, threads free the objects using the reclamation technique's synchronization.
In the context of reclamation algorithms, a thread with a reference to a shared object in the data structure is termed a \textit{reader}, while a thread deleting an object from the data structure is called a \textit{reclaimer}. Readers can save a pointer or a timestamp, representing a collection of objects, in their local or at shared memory locations. These are referred to as \textit{reservations}.
These \textit{reservations} are made globally visible for reclaimers, referred to as \textit{publishing}.
Threads are considered to be in a \textit{quiescent state} between consecutive data structure operations, meaning they do not access any shared objects. 

% \subsection{Object States}

% \end{enumerate}

% \textbf{-- shall we add popHE code here or in appendix?}

The synchronization mechanisms, vary between techniques, and ensure that the reclaimers only free \textit{safe} objects and \textit{readers} do not access \textit{unsafe} objects. For example, in hazard pointers, synchronization happens by requiring \textit{readers} to reserve and publish objects before accessing them, and \textit{reclaimers} to first scan reservations to identify all the \textit{unsafe} objects, and subsequently freeing only the \textit{safe} objects from their retire list.
Similarly, in epoch based techniques, \textit{reclaimers} must wait before freeing an object in their retire list until all threads have gone quiescent at least once since the object was retired. 
% The aforementioned condition ensures that by going quiescent, threads loose all reference to data structure objects and acquire new reference by traversing from the entry point again with a new data structure operation. 

% In this section, we revisit hazard pointers (HP)~\cite{michael2004hazard} and a read-copy-update style (RCU)~\cite{hart2007performance} implementation of epoch based reclamation, hereafter, referred to as epoch based reclamation (EBR) unless specified otherwise.
% HP is actively utilized in MongoDB and at Facebook, while the RCU is part of the Linux OS. 
% Both techniques, have recently been proposed for addition to C++ standard  library~\cite{michaelKenny2017proposed,michael2017hazard}.

This section revisits hazard pointers (HP)~\cite{michael2004hazard} and read-copy-update (RCU) style~\cite{hart2007performance} implementations of epoch based reclamation (EBR). HP is used in MongoDB and in Folly- Facebook's open source C++ library, and RCU is part of the Linux OS. Both techniques have been proposed for the C++ standard library~\cite{michaelKenny2017proposed,michael2017hazard}.
Therefore, we have chosen these two techniques for discussion in the background section and will later build upon them to explain our publish on ping implementations of these algorithms.
% We discuss these techniques in the background section and will later explain our publish on ping implementations.

\subsection{Hazard Pointers}
\subsubsection{The Overview}
The key principle underlying the operations of Hazard Pointers (HP) involves a contract between \textit{readers} and \textit{reclaimers}. \textit{Readers} \textit{reserve} and \textit{publish} pointers to objects currently being accessed at single-writer multi-reader (SWMR) locations for \textit{reclaimers}. 
\textit{Reclaimers}, in turn, ensure they scan all the SWMR locations to collect all reserved objects and only free those whose reservations have not been \textit{published}. This ensures safety from use-after-free errors.

Crucial to publishing of reservations in HP is that every read of a shared object pointer should execute a memory fence. Specifically, \textit{readers} should follow these steps to timely publish their reservation to a pointer: 1) save the pointer to the object being read, 2) execute a memory fence, 3) validate that the pointer saved in the step 1 is still \textit{reachable}, if not, retry or abort the operation.
This ensures that the pointer was \textit{reachable} at the time it was reserved (in step 1) \textendash a crucial condition for correct application of HPs~\cite{michael2004hazard}.
Without the memory fence, the reservation of the pointer in step 1 could be reordered after the validation of reachability step (step 3). This could lead to reservation of an object which has already been deleted, resulting in unsafe accesses in the future.
% [\textcolor{red}{an example is better?}]. 

\subsubsection{The Problem}
% The Hazard Pointers is robust (or lock-free) because at a given time during an execution only $O(N^2*H)$ objects may not be reclaimed, where $N$ is the number of threads and $H$ is the number of hazard pointers a thread can hold at any given time. In other words, a delayed thread will only prevent a bounded number of objects from getting reclaimed thus it doesn't impact reclamation of other threads, making them attractive in practise.
The per read memory fences incur high overhead, leading to poor scaling of data structures. Our Perf tool analysis, on a Hazard-Michael list of size 100 nodes where 128 threads execute 50\% inserts and 50\% delete operations showed that searches approximately spend $\approx$ 50\% of CPU cycles on reading HPs, whereas searches in a leaky implementation of the same list only spends $approx$ 15\% of CPU cycles in reading the HPs.  

Several subsequent techniques have aimed to reduce or eliminate the need for memory fence during traversals in HP.
For example, Cadence~\cite{balmau2016fast} utilizes system level memory fences triggered by context switches. 
This allows \textit{readers} to defer publishing reservations (using explicit memory fence) until a context switch occurs. \textit{Reclaimers}, to free a retired object, wait for a context switch to occur since the object was retired. This waiting period ensures that the reservations to the object, if any, become visible and the \textit{reclaimer} can free it if it is not found to be reserved.

Cadence employs auxiliary threads pinned to each core to force the context switches at regular intervals. These auxiliary threads compete with primary worker threads that execute data structure operations, impacting performance. Moreover, the \textbf{\textit{assumption}} that rescheduling always triggers a memory fence is architecture dependent~\cite{mckenney2010memory, balmau2016fast}.
This mechanism requires the instrumentation of data structure objects with a timestamp to determine whether a global memory fence has occurred since the object was retired, affecting the memory layout of the underlying data structure. 
Additionally, the overhead due to auxiliary threads enforcing a global memory fence, though amortized over multiple operations, is incurred even if threads do not reclaim. So, the issue of uneven overhead between readers and reclaimers is still present.

% alleviate the cost of the read operations in Hazard Pointers. For example, Cadence in \cite{balmau2016fast} attempts to reduce the cost of frequent memory fences by leveraging an extra process pinned to hardware cores (called rooster process) along with the working processes that execute data structure operations.
% These rooster processes periodically force the working threads to get rescheduled out of its core. Assuming the rescheduling always causes a memory fence leading to reservations becoming globally visible. 
% \textcolor{red}{Should I point out Safety concerns with rooster processes?}  
Dice et al. in ~\cite{dice2016fast} discuss a technique to eliminate memory fences from traversals by leveraging write-protect feature in modern operating systems, which enforces a global memory barrier.
\textit{Reclaimers} briefly write-protect all memory pages associated with hazard pointers and then remove the protection, ensuring a global memory fence is executed. This makes all the hazard pointers visible before attempting to free retired objects.
However, this technique requires all threads to block in an interrupt handler until the write protection is revoked. In the same paper, the authors also discuss a hardware extension suggesting the addition of a dedicated store buffer unit along with two new instructions to maintain hazard pointers.

Another technique, Hazard Eras (HE)~\cite{ramalhete2017brief}, which emerged after the former two, attempts to amortize the cost of the per-read memory fences by using timestamps. \textit{Readers} reserve current eras, instead of pointers, represented by monotonically increasing global timestamps, whenever they read a pointer. The memory fences to timely publish the reserved eras are only incurred when the epoch changes concurrently with a read of an object pointer.    
In more detail, Hazard Eras leverages infrequent changes in epochs to avoid incurring memory fences at every read.
\textit{Reclaimer} free only those nodes whose lifetime does not intersect with epochs reserved by all threads. Although the amount of garbage is bounded but it still can be very high proportional to the all the memory whose lifetime intersects with a reserved era, and is still slower in many workloads as observed in our experiments.

% However, the frequency of the memory fences is reduced at the expense of a slow rate of change of the global timestamp. Reducing the rate of change of the global timestamp decreases the number of memory fences but increases the memory footprint, and vice versa.

% --
% HE, on the other hand, reserves more precise epochs. Essentially, it is like hazard
% pointers, but instead of reserving each pointer, it reserves the global epoch at the time
% it accesses a node. The reclaiming thread frees only those nodes whose lifetime does not
% intersect with epochs reserved by all threads
% --

The process-wide memory barrier \texttt{sys\_membarrier}\cite{membarrierSystemwide} on Linux reduces the asymmetric overhead between readers and reclaimers and is used in both the RCU in the kernel and Folly's HP implementation. However, the availability and implementation of \texttt{sys\_membarrier} (which can be blocking) vary across kernel versions and architectures. When unavailable, the system may fall back to \texttt{mprotect}, which can degrade performance\cite{mprotectoverhead}. 
Additionally, \texttt{sys\_membarrier} has been linked to security vulnerabilities~\cite{CVE202426602}. While HP with \texttt{sys\_membarrier} offers significant improvement over the original algorithm, it can still be 12\%-40\% slow, as our experiments will show.

% The process-wide memory barrier-\texttt{sys\_membarrier}~\cite{membarrierSystemwide} on Linux helps reduce the asymmetric overhead between readers and reclaimers. Infact, it is used for RCU in kernel and Folly's HP implementation.
% However, the availability of \texttt{sys\_membarrier} and its specific implementation (can be blocking) varies between different kernel versions and architectures. In case the \texttt{sys\_membarrier} is not available, executions may fall back to \texttt{mprotect} system call that could degrade performance~\cite{mprotectoverhead}. 
% In addition, use of \texttt{sys\_membarrier} has been the cause of security vulnerability~\cite{CVE202426602}. The HP implementation with \texttt{sys\_membarrier} though improves significantly over the original algorithm but is still 12\%-50\% slower than the proposed mechanism, as we will see in our experiments.

% \textcolor{red}{--timestamps or epoch? I mostly used epoch but here timestamp feels good .. I am not consistent... delegate decision to  next pass.}

% \textcolor{red}{--We call HE, HP , IBR pointer based as they do per pointer read some actions. Shoudl I generalize the fence overhead to HE, IBR? I do have experiments with IBR}

\subsection{Epoch Based Reclamation}
\label{sec:rcu}

\subsubsection{The Overview}
Epoch based reclamation (EBR) has multiple variants, such as those proposed by Harris ~\cite{harris2001pragmatic}, Fraser \cite{fraser2004practical}, RCU~\cite{hart2007performance}, and Brown's DEBRA~\cite{brown2015reclaiming}.
The key concept across these algorithms is threads execute data structure operations in sequence of monotonically increasing timestamps called  epochs. \textit{Readers} publish the current epoch they are executing in and \textit{reclaimers} can free objects retired in or before a given epoch $e$ once every thread has completed execution in epoch $e$ or earlier and has transitioned to a more recent epoch (greater than $e$). 
Implying that for objects retired on or before epoch $e$, all threads have gone quiescent at least once, making the objects \textit{safe} for reclamation. In RCU~\cite{michaelKenny2017proposed} terminology, while executing data structure operations, threads are assumed to be executing a read-side critical section. A \textit{reclaimer} can only free objects that are \textit{retired} before the beginning of the oldest read-side critical section. 
Pseudocode appears in the appendix \secref{apxrcu}.

% \algoref{rcuebr}, shows an example implementation.
% It maintains a monotonically increasing shared \func{epoch} variable (assuming it never overflows). It is incremented after a certain number of operations to represent a progression of epochs. Threads, utilize a \func{reservedEpoch} array, with one single-writer multi-reader slot for each thread to save and publish the epoch they are executing in. A thread declares its entry into a read-side critical section by publishing the current value of \func{epoch} in its \func{reservedEpoch} slot. It declares the exit from the previous read-side critical section by publishing a maximum possible value MAX (assuming \func{epoch} can never be the same as this value).

% When a thread decides to reclaim its \func{retireList}, it finds the minimum epoch reserved by a thread from the \func{reservedEpoch} array. It then frees those objects retired before the minimum reserved epoch. In the implementation, this is identified by comparing the \func{retireEpoch} of the retired objects which are associated with the objects at the time it is appended to its \func{retireList}.
% Since frequent increment of the shared \func{epoch} and freeing of \func{retireList} is an overhead, these are amortized over multiple operations to boost performance.

\subsubsection{The Problem}
\label{sec:rcurobustness}
The main concern with EBR is that a thread could get stuck in a data structure operation, leading to a delayed exit from an old read-side critical section due to arbitrary system level reasons, such as page fault servicing, thread scheduling etc. These delays might be significant enough to cause the minimum declared epoch to lag far behind the current global \func{epoch}.
This situation prevents all threads from freeing their retire lists, resulting in a drastic increase in system memory consumption and possibly leading to out-of-memory errors. This issue is commonly referred to as the lack of \textit{robustness}. 

% For example, in the given implementation, a thread stuck in a long-running data structure operation, after reserving a small epoch value, will impede other threads from preventing all objects retired on and after the reserved epoch, even though all other threads are at the much larger, more recent epochs. Consequently, \func{retireList} of all threads could continuously grow, leading to increasingly higher memory consumption.

% Various other techniques like NBR, VBR and concurrently with this work HP-BRCU achieve robustness and efficiency but compromise on programmability of original HPs to varying degrees (which is inline with ERA theorem\cite{sheffi2023era}).
% Common to these techniques is that they require programmers to install checkpoints (one per read phase in NBR and multiple per read phase in VBR and HPBRCU) and assume that restarting a thread executing at an arbitrary region in code can abort and restart from a previous checkpoint.

% In addition in NBR and HPBRCU programmers must identify regions in code which are \textit{unsafe} for signal-induced restarts, eg regions in code where treads might have taken locks or have done partial updates~\cite{david2015asynchronized, heller2005lazy, harris2001pragmatic}.
% Moreover, beforehand identify and reserve all the nodes that could be accessed within such regions.
% This can be difﬁcult for arbitrary data structures\cite{Cohen_thesis16, sheffi2021vbr, kim2024expediting}.
% These restarts could be overhead in varying degrees. 

Various techniques, such as NBR~\cite{singh2021nbr, singhTPDS2023NBRP}, VBR~\cite{sheffi2021vbr}, and HP-BRCU~\cite{kim2024expediting}, achieve robustness and efficiency but compromise the programmability of original HPs to varying degrees, in line with the ERA theorem \cite{sheffi2023era}. These methods typically require programmers to install checkpoints (one per read phase in NBR, and multiple per read phase in VBR and HP-BRCU) and assume that a thread can abort and restart from a previous checkpoint while it is executing in an arbitrary code region.

Additionally, in NBR and HP-BRCU, programmers must identify regions in code that are \textit{unsafe} for signal-induced restarts, such as those where threads might have taken locks or performed partial updates \cite{david2015asynchronized, heller2005lazy, harris2001pragmatic}. They must also identify and reserve all nodes that could be accessed within these regions beforehand. This process can be challenging for arbitrary data structures \cite{Cohen_thesis16, sheffi2021vbr, kim2024expediting}. These restarts can introduce varying degrees of overhead and non-trivial reasoning about the correctness of the memory reclamation scheme for practitioners.

% add complexity in comparison to original approaches.
% Ours is Similar to HP incomplexity and applies to allDS to which HP applies.
% --\textcolor{red}{restarting could be non trivial in many data structures: n. Devising recovery
% code requires non-trivial reasoning about the correctness of the
% memory reclamation scheme.}

\subsection{Summary: The Problem and The Solution}
Programmers face unattractive choices: choose robust but slow hazard pointers\cite{michael2004hazard}, opt for robust and fast ad hoc era or pointer reservation-based techniques~\cite{balmau2016fast, dice2016fast, ramalhete2017brief}, but tolerate intrusiveness or less portable hardware dependent solutions, or opt for a fast but not robust EBR\cite{brown2015reclaiming, harris2001pragmatic, fraser2004practical, hart2007performance} or opt for hybrid techniques that are fast and robust but tradeoff programmability in varying degrees~\cite{kim2024expediting, sheffi2021vbr, singh2021nbr, singhTPDS2023NBRP}.  

Consequently, it will be desirable to have a simpler solution that is fast and robust and yet maintains a similar programmability property like the original hazard pointers. Hazard eras and \texttt{sys\_membarrier} was on right path for achieving this but it seems to be slow, as we will show in our experiments.

This paper introduces a fast and non-intrusive publish-on-ping paradigm that uses signals without inducing complexity of signal-based restarts like the earlier techniques to accelerate hazard pointers. 
We apply publish-on-ping to hazard pointers and hazard eras that result in the new algorithms HazardPtrPOP and HazardEraPop, respectively.
Building upon HazardPtrPOP, we implement a variant of hazard pointers using EBR to accelerate hazard pointers in the common case where threads are not frequently delayed. In rare cases where thread delays are suspected, \textit{reclaimers} can continue freeing their retire list using publish-on-ping. The technique is unique in the sense that it does not require switching between the two reclamation schemes and two threads could be reclaiming at the same time in either common EBR or less common HazardPtrPOP way.   
These techniques are backward compatible with hazard pointers and hazard eras, retaining their ease of programmability and apply to all data structures hazard pointer applies to.

\section{Publish On Ping}
\label{sec:popoverview}
\subsection{The Overview}
% The key aspect of our technique is that threads can read new pointers, reserve them locally without immediate publishing to \textit{reclaimers}. This eliminates the need for costly memory fences per read. \textit{Reservations} are published only when some thread attempts to reclaim its retire list. 
% In other words, when a thread wants to reclaim its retire list, it signals all participating threads to publish their local \textit{reservations} to shared locations. 
% The \textit{reclaimer} then scans these shared locations to collect all the \textit{reservations}. Subsequently, the \textit{reclaimer} frees all the retired objects in its retire list that are not reserved by other threads.

% The semantic publish-on-ping (POP) behavior is implemented with the help of a POSIX signal and a corresponding signal handler. \textit{Reclaimers} use a \func{pthread\_kill} call to signal all other threads in the system (\textbf{ping}). All other threads execute a signal handler to assign their local reservations to corresponding shared locations (\textbf{publish}).

% In a nutshell, POP eliminates the memory fence overhead from the principal traversal path by allowing threads to only publish the reservations when demanded by infrequent reclamation events through a simple and neat use of signalling, i.e without forcing thread restarts. 
% This elimination of the overhead from the read path benefits data structures, in practice, where read-dominated workloads are common. 

A key aspect of our technique is that threads can read new pointers and reserve them locally without immediately publishing these reservations to \textit{reclaimers}. 
This approach eliminates the need for costly memory fences on every read. \textit{Reservations} are published only when a thread attempts to reclaim its retire list. Specifically, when a thread wants to reclaim its retire list, it signals all participating threads to publish their local \textit{reservations} to shared locations. The \textit{reclaimer} then scans these shared locations to collect all the \textit{reservations} and subsequently frees all retired objects in its retire list that are not reserved by other threads.

The publish-on-ping (POP) behavior is implemented using a POSIX signal and a corresponding signal handler.
\textit{Reclaimers} use the \func{pthread\_kill} call to signal all other threads in the system (\textbf{ping}). The other threads execute a signal handler to assign their local reservations to the corresponding shared locations (\textbf{publish}) for reclaimers.

% \textit{Reclaimers} use the \texttt{pthread\_kill} call to signal all other threads in the system (\textbf{ping}). The other threads then execute a signal handler to \textbf{publish} their local reservations to the corresponding shared locations.

In summary, POP removes the memory fence overhead from the main traversal path by having threads publish reservations only when reclamation events occur, with a simple signaling mechanism. This avoids forcing thread restarts (unlike previous signalling based techniques, such as NBR(+), DEBRA, and PEBR) and reduces overhead on the read path, which is particularly beneficial for data structures with read-dominated workloads.
Moreover, POP is equally effective for other related reservation based techniques, incurring similar per-read memory fences during traversals, such as, Hazard Eras (HE).
We apply POP to HE and observe performance improvements, as demonstrated in \secref{experimentalEvaluation}.
% HE  though amortizes the publishing of era reservations but does not remove it completely from the traversal (or read) path.

% \textcolor{red}{[I implemented pop Interval Based reclamation too but haven't reported here. Dilemma, should I include it, mightgo in appendix? May be, I should leave those for sake of focussed manuscript. Delegated to next pass.]}

\ignore{
One important challenge, however, that needs to be addressed is that \textit{reclaimers}, after sending pings to all threads, should provably establish a time when all \textit{reservations} are considered published. This is crucial to safely free objects in the retire lists of the objects. We discuss this aspect in detail while describing our example HazardPtrPOP implementation, in the next section. 
}
\section{Algorithms}
\label{sec:algorithms}

\subsection{HazardPtrPOP}
Having discussed publish-on-ping, we now apply it to HP to eliminate the need for publishing reservations during traversals or reads of new objects in data structures. The resulting algorithm is termed Hazard Pointer Publish-on-Ping (HazardPtrPOP). 
First, we provide an overview of a typical HP implementation.

\subsubsection{Programmer's view of HP}
For programmers, a standard Hazard Pointers reclamation interface includes the following three main functions: 

(1) \Call{read}{ }: Used for every read of a new data structure object. \textit{Readers} use \Call{read}{} to ensure that threads \textit{reserve} and \textit{publish} the object to a shared location.
(2) \Call{clear}{ }: Employed to remove reservations when threads finish accessing an object or exit the operation. 
(3) \Call{retire}{ }: Used during update operations, \textit{reclaimers} employ \Call{retire}{} to append \textit{deleted} objects to their retire lists. During the \Call{retire}{}, \textit{reclaimers} collect every other threads's published reservations. Subsequently, in an iterative fashion, they free those objects from their retire list that were not present in their collected reservations.

% Threads maintain a single writer multi reader list of hazard pointers which is used to protect objects they are currently accessing from being freed by a reclaiming thread. Threads also maintain a per thread retire list where all thread collect the retired objects. If the retire list reaches to a threshold size, then threads collect all reservations followed by freeing all the objects that were not reserved. Eventually, when a thread completes its data structure operation it executes clear() which removes the reserved objects from its reservation list so that the reclaiming threads do not skip the objects from freeing even though it is no longer being accessed by the thread that reserved them.

\begin{algorithm}[t]
\footnotesize
    \caption{HazardPtrPOP: Hazard Pointer Publish-on-Ping.}
    \label{algo:pophp}
    \begin{algorithmic}[1]
    % \Statex \textbf{global:}
    \State \texttt{const int reclaimFreq} \Comment{frequency of reclaiming retire list} \label{lin:var-reclaimfreq}
    % \Statex \textbf{per thread local structures:}
    \State \texttt{thread\_local int tid} \Comment{current thread id}
    \State \texttt{list<T*> retireList [NTHREAD]} \label{lin:var-retirelist}
    % \Statex
    % \Statex \textbf{per thread shared structures:}
    \BeginBox[draw=black, dashed]
        \State \texttt{T* localReservations [NTHREAD][MAX\_HP]} \label{lin:var-localreservations}
        \State \texttt{atomic<T*> sharedReservations [NTHREAD][MAX\_HP]} \label{lin:var-sharedreservations} 
        \State \texttt{atomic<int> publishCounter [NTHREAD]} \label{lin:var-publishcounter} 
        \State \texttt{thread\_local int collectedPublishCounters [NTHREAD]} \label{lin:var-collectedpublishcounters} 
    \EndBox
        \Procedure{\texttt{T*} read}{\texttt{atomic<T*> \&ptrAddr, int slot}} \label{lin:proc-hppopread}
            \Repeat
                \State \texttt{T* readPtr $\gets$ *ptrAddr}
                \BeginBox[draw=black, dashed]
                \State \texttt{localReservations[tid][slot] $\gets$ readPtr} \LComment{no store load fence needed.}
                \EndBox
            \Until{\texttt{readPtr $=$ *ptrAddr}} \label{lin:loopexit}
            \State \Return readPtr \label{lin:return}
        \EndProcedure
    \Statex
        \Procedure{retire}{\texttt{T* ptr}} \label{lin:proc-retire}
            \State \texttt{myRetireList $\gets$ retireList[tid]}
            \State \texttt{myRetireList.append(ptr)}
            \If {\texttt{myRetireList.size() $\geq$ reclaimFreq}}
                \BeginBox[draw=black, dashed]
                    \State {\Call{collectPublishedCounters}{ }} \label{lin:collectpublishedcounters}
                    \State {\Call{\textbf{pingAllToPublish}}{ } } \label{lin:pingalltopublish}
                    \State {\Call{waitForAllPublished}{ }} \label{lin:waitforallpublished}
                \EndBox
                % \State {\Call{collectAllReservations}{ }}
                \State {\Call{reclaimHPFreeable}{myRetireList}} \label{lin:reclaimHPFreeable}
            \EndIf
        \EndProcedure    
    \Statex
        \Procedure{clear}{ } \label{lin:proc-clearall}
            \For{$slot=0, \dots, MAX\_HP$}
                \BeginBox[draw=black, dashed]
                    \State \texttt{localReservations[tid][slot] $\gets$ NULL}\label{lin:setnull}
                    % \LComment{no store load fence needed}
                \EndBox
            \EndFor        
        \EndProcedure        
\algstore{algpophp}
\end{algorithmic}
\end{algorithm}

\begin{algorithm}[t] 
\footnotesize
\caption{HazardPtrPOP: Continued.}
\label{algo:pophpcontd}
\begin{algorithmic} [1]                   % enter the algorithmic environment
\algrestore{algpophp}
        \Procedure{reclaimHPFreeable}{myRetireList} \label{lin:proc-reclaimhpfreeable}
                \LComment{collect all published reservations}
                \State \texttt{set<T*> collectedReservations $\gets$ \{\}} \label{lin:collectstart}
                % \ForAll {\texttt{tid}}
                    \ForAll {\texttt{<tid, slot> $\in$ sharedReservations[tid]}}
                        \State {\texttt{objPtr $\gets$ sharedReservations[tid][slot]}}
                        \State {\texttt{collectedReservations.insert(objPtr)}}
                    \EndFor \label{lin:collectend}
                % \EndFor                
                \LComment{free all objects not reserved}
                \ForAll {\texttt{objPtr $\in$ myRetireList}} \label{lin:reclaimstart}
                    \If{\texttt{objPtr $\notin$ collectedReservations}}
                        \State {\texttt{free(objPtr)}}
                    \EndIf
                \EndFor \label{lin:reclaimend}
        \EndProcedure

\BeginBox[draw=black, dashed]
    \Statex
        \Procedure{pingAllToPublish}{ } \label{lin:proc-pingalltopublish}
            \ForAll {\texttt{othertid $\neq$ tid} }
                \State {\texttt{pthread\_kill(othertid, \dots)}}
            \EndFor
        \EndProcedure
    % \Statex
        \LComment{signal handler}
        \Procedure{publishReservations}{ } \label{lin:proc-publishreservations}
            % \State \Call{\textbf{publishReservation}}{\texttt{tid}}
            \For{$ihp=0, \dots, MAX\_HP$}
                \State \texttt{sharedReservations[tid][ihp] $\gets$ localReservations[tid][ihp]}
                % \LComment{store load fence needed. TBD : ONLY 1 FENCE?}
            \EndFor
            \BeginBox[draw=blue, dotted]
            \State \texttt{publishCounter[tid] $\gets$ publishCounter[tid]+1} \label{lin:publishcounter}
            % \Comment{TBD: is fence not implicit or clear to readers?}
            \EndBox
        \EndProcedure

    \Statex
        \Procedure{collectPublishedCounters}{ } \label{lin:proc-collectpublishedcounters}
            \ForAll {\texttt{tid}}
                \State {\texttt{collectedPublishCounters[tid] $\gets$ publishCounter[tid]}}
            \EndFor
        \EndProcedure    
    \Statex
        % \LComment{Completes in constant time as signals are delivered immediately~\cite{singhTPDS2023NBRP, kim2024expediting, brown2015reclaiming} }
        \Procedure{WaitForAllPublished}{ } \label{lin:proc-WaitForAllPublished}
            \LComment{establish all threads have executed signal handler}
            % \ForAll {\texttt{tid}} \label{lin:verifstart}
            % \Repeat
            %     \State {\texttt{collectedPublishingCounter[tid] $\gets$ publishCounter[tid]}}
            % \Until{\texttt{collectedPublishingCounter[tid]+1 $\ge$ publishCounter[tid]}}
            \ForAll {\texttt{othertid $\neq$ tid}} \label{lin:verifstart}
            \Repeat
                % \State {\texttt{collectedPublishingCounter[tid] $\gets$ publishCounter[tid]}}
            % \Until{\texttt{collectedPublishingCounter[tid]+1 $\ge$ publishCounter[tid]}}
            \Until{\texttt{publishCounter[othertid] $>$ collectedPublishCounter[othertid]}}
            \EndFor \label{lin:verifend}
        \EndProcedure    
\EndBox
    \end{algorithmic}
\end{algorithm}

\algoref{pophp} and \algoref{pophpcontd} show the implementation of the proposed HazardPtrPOP algorithm.
From a programmer's perspective, it maintains the same interface as in HP, making it backward compatible with HP.
Similarly to HP, threads maintain a list per thread, depicted as \func{retireList} (\lineref{var-retirelist}) to which they append the retired objects with call to \Call{retire}{ } within their update operations.

Contrary to the publish eagerly paradigm in HP, threads in HazardPtrPOP locally save the objects at their own list of slots in \func{localReservations} (\lineref{var-localreservations}) array during \Call{read}{}. 
The \Call{read}{} (\lineref{proc-hppopread}) procedure repeatedly reads the pointer to the object, saves it in a corresponding slot in \func{localReservations}, and then rereads the pointer.
Assuming a system with maximum of \func{NTHREADS} threads, every thread has the same fixed number of slots represented by \func{MAX\_HP}. 
The loop exits only when the pointer remains unchanged on the second read, upon which the pointer is returned.
When a thread is about to go \textit{quiescent}, i.e., while exiting the data structure operation, it resets the local reservations by setting the corresponding slots to NULL (\lineref{setnull}) by using \Call{clear}{}. This allows the reserved nodes to get freed when they are not in use.
These local reservations are published to a corresponding shared array of single-writer multi-reader slots, called \func{sharedReservations} (\lineref{var-sharedreservations}).

% During \Call{retire}{} (\lineref{proc-retire}), a \textit{reclaimer} invokes \Call{pingAllToPublish}{ } to trigger the publishing of local reservations when the size of its \func{retireList} exceeds a given threshold set in \func{reclaimFreq} (\lineref{var-reclaimfreq}).
% In \Call{pingAllToPublish}{} (\lineref{proc-pingalltopublish}) a \textit{reclaimer} uses \func{pthread\_kill()} to ping all threads.
% The threads receiving these pings execute a signal handler, called \Call{publishReservations}{} (\lineref{proc-publishreservations}). Within the signal handler, other threads write all their local reservations to the shared array.
% Once all threads complete execution of the \Call{publishReservations}{}, then the reservations become visible to the \textit{reclaimer}.
% After every thread publishes its reservations, the \textit{reclaimer} can invoke \Call{reclaimHPFreeable}{} (\lineref{reclaimHPFreeable}) to 
% collect all the reservations and frees those that are not in the collected reservations (\lineref{collectstart}-\lineref{reclaimend}), similar to HP.

In the \Call{retire}{} procedure (\lineref{proc-retire}), the \textit{reclaimer} triggers the publishing of local reservations by calling \Call{pingAllToPublish}{} when the size of its \func{retireList} exceeds the threshold set with \func{reclaimFreq} (\lineref{var-reclaimfreq}). Within \Call{pingAllToPublish}{} (\lineref{proc-pingalltopublish}), the \textit{reclaimer} employs \func{pthread\_kill()} to send pings to all threads. The threads that receive these pings execute a signal handler, named \Call{publishReservations}{}, (\lineref{proc-publishreservations}), which writes all their local reservations to the shared array. Once all threads have completed \Call{publishReservations}{}, the reservations become visible to the \textit{reclaimer}. After all threads' reservations are published, the \textit{reclaimer} can then call \Call{reclaimHPFreeable}{} (\lineref{reclaimHPFreeable}), which similar to HP gathers up all the reservations and frees those that are not part of the collected reservations (\lineref{collectstart}-\lineref{reclaimend}).

% free all the retired objects, provided they are not reserved (\lineref{proc-reclaimhpfreeable}). Precisely, within \Call{reclaimHPFreeable}{ }, the \textit{reclaimer}

% These local reservations are only published on demand globally when a thread is about to reclaim objects in its retireList. Specifically, a reclaiming thread when exceeds a certain size of its retireList, signals all threads using pingToPublish() to publish their reservations in its single writer multi reader slots in sharedReservations.

% pingToPublish(), essentially invokes pthread\_killl to signal (or ping) all threads. This causes all threads to execute the signal handler publishReservations(). It simply copies the reserved objects in the slots at the localReservations to the single writer multi reader slots in the sharedReservations. Once all threads complete execution of the publishReservations(), the reservations become visible to the reclaimers.

% An important piece for the safety of reclamation is the challenge of how do we decide a time when it can be said that all threads have executed their \Call{publishReservations}{ } so that the \textit{reclaimer} could safely proceed to free its \func{retireList} after the call to \Call{pingAllToPublish}{ } returns.

A key factor in ensuring safe reclamation (avoiding use-after-free) is determining the point at which we can confirm that all threads have completed their \Call{publishReservations}{ } calls, allowing the \textit{reclaimer} to safely free its \func{retireList} once the \Call{pingAllToPublish}{ } function returns.
To establish such a time, HazardPtrPOP uses \func{publishCounter} array (\lineref{var-publishcounter}) where each slot is a monotonically increasing counter assigned to a specific thread. 
Each thread in the system increments its slot in the \func{publishCounter} (\lineref{publishcounter}) after finishing publishing.
% The \textit{reclaimers} observe every thread's \func{publishCounter} value before and after pinging all threads to assert that every thread has finished publishing, based on the comparison between the previously read \func{publishCounter} values and the reread values. 
% Precisely, the \textit{reclaimers} read every thread's \func{publishCounter} value in to their thread local \func{collectedPublishCounters} array using \Call{collectPublishedCounters}{ } at \lineref{collectpublishedcounters}. Then, they invoke \Call{pingAllToPublish}{ } (\lineref{pingalltopublish}), followed by call to \Call{waitForAllToPublished}{ } (\lineref{waitforallpublished}), wherein, it repeatedly rereads every thread's \func{publishCounter} value and compares it with the previously observed values and only exits the loop when all threads have incremented their \func{publishCounter} value at least once since the time the \textit{reclaimer} collected the \func{publishCounter} values (\lineref{verifstart} - \lineref{verifend}).

A \textit{reclaimer} monitor each thread's \func{publishCounter} value before and after pinging all threads to assert that every thread has completed publishing, based on comparing the previously read \func{publishCounter} values with the reread values.
Specifically, the \textit{reclaimer} read each thread's \func{publishCounter} value into their thread-local \func{collectedPublishCounters} array using \Call{collectPublishedCounters}{ } at \lineref{collectpublishedcounters}. Then, it calls \Call{pingAllToPublish}{ } (\lineref{pingalltopublish}), followed by \Call{waitForAllToPublished}{ } (\lineref{waitforallpublished}), during which the reclaimer repeatedly reread each thread's \func{publishCounter} value and compare it with the previously recorded values, only exiting the loop when all threads have incremented their \func{publishCounter} value at least once since the \textit{reclaimer} collected the \func{publishCounter} values (\lineref{verifstart} - \lineref{verifend}).
(When multiple reclaimers send signals simultaneously, the signals are effectively coalesced,  and a reader publishing reservations once is sufficient to satisfy all concurrent reclaimers interested in knowing about those reservations.)
% This ensures that  and a reader publishing reservations once is sufficient to satisfy all concurrent reclaimers interested in knowing about those reservations.

% An important point we deliberately delayed discussing for the sake of clarity is... it seem tough to ensure the time which is safe to start reclaiming... in theory waiting may appear to kill progress
\subsubsection{Limitations of POP and signals}
%
%We assume \Call{waitForAllToPublished}{} completes in bounded time.
The POP mechanism assumes that, upon being signaled, all threads will complete executing their signal handlers within a bounded time. This assumption is crucial for ensuring that all threads publish their reservations within a finite number of steps, allowing the \textit{reclaimer} to eventually exit its \Call{waitForAllToPublished}{} loop.
In practice, signal delivery occurs in bounded time, as demonstrated by prior signal-based techniques \cite{brown2015reclaiming, kang2020marriage, alistarh2018threadscan, alistarh2017forkscan, singh2021nbr} and verified by studies on the timeliness of signal delivery \cite{singhTPDS2023NBRP} on modern architectures.
Note that signal delivery timings are expected to get an order of magnitude faster with user space IPIs~\cite{useripilinux}.

The key advantage of using POSIX signals is that we can relax the traditional thread failure model in the asynchronous shared-memory setting.
%where one cannot distinguish between a slow thread and a crashed thread.
In theory, a thread can run arbitrarily slowly or halt altogether, and one cannot distinguish between a slow thread and a halted one.
In practice, threads that appear to be delayed are either busy with other work, trapped in infinite loops, or descheduled.
The operating system knows which threads are running, descheduled, or have terminated or become zombies (awaiting events to terminate).

The \textit{pthread\_kill} function used to signal (ping) threads returns an error code if a thread is a zombie, or terminated, allowing a reclaimer to ignore such threads.
This leaves running threads and descheduled threads.
%Threads that are actively running on a CPU typically execute instructions at similar rates.
Running threads will be interrupted by a signal, and quickly publish their reservations.
As for descheduled threads, modern schedulers are (somewhat) fair, and ensure each thread is scheduled to run periodically, at which point a thread will publish its reservations.

The worst case scenario for POP is an \textit{oversubscribed} system with many more threads than CPUs, as a reclaimer will need to wait for all threads to be scheduled to publish reservations.
However, concurrent data structures tend to scale negatively as the number of threads grows beyond the number of CPUs, so this scenario is not our focus.
That said, our experiments include oversubscription, and POP performs surprisingly well despite moderate oversubscription. %we do not focus on this scenario is not a priority for us.

\subsubsection{Correctness and progress}

\begin{assumption}
\label{asm:sigasm}
Threads publish their reservations in a bounded time after being pinged.    
\end{assumption}

\begin{property}
    (Safety) HazardPtrPOP avoids use-after-free errors. 
\end{property}
In order to prove HazardPtrPOP is safe, we need to establish that any \textit{reclaimer} will not free an object which other threads could subsequently access.

$Wlog.$, by the way of contradiction, let us assume, a thread $T1$ frees an object $o$ at a time $t1$ which is subsequently accessed by another thread $T2$ at a later time $t2$ ($t1 < t2$). 
In order to access $o$, $T2$ must successfully protect it at an earlier time $t2'$, such that $t2' < t2$. 
Similarly, to free $o$, $T1$ retires, then pings all threads to publish their reservations, and then for a bounded time waits to ensue that all threads complete publishing their reservations at a time $t1'$, such that $t1' < t1$.
Now, Two cases arise. 
First, $t1' < t2'$, i.e., $T2$ published all its reservations for $T1$ before it reserved $o$ at $t2'$.
In this case, since $n$ was already retired by $T1$, $T2$ will fail validation while reserving $o$. 
Thus, $T2$ can not access $o$ without successfully reserving (HP requirement that threads reserve before access).
In the second case, $t2' < t1'$, i.e $o$ was successfully reserved before $T1$ requested $T2$ to publish the reservations.
Note, in this case, $T2$ would have published the reservation which $T1$ will scan and skip freeing $n$ as it is guaranteed to find it in the reservation list of $T2$ (\asmref{sigasm}). Hence, HazardPtrPOP is immune to use-after-free errors.  

\begin{property}
    (Liveness) HazardPtrPOP is robust.
\end{property}
A thread accumulates $r$ nodes in its retire list of which $N\times H$ nodes may be reserved, where $N$ is the number of threads and $H$ is the maximum number of reservations that $N$ threads could hold at a given instance in time. This implies, at a given time, a thread could free at least $r - N\times H$ nodes and at most $N\times H$ nodes may not be freed.
Since $N\times H$ is a constant, the amount of unreclaimed garbage per thread is always constant. Therefore, popHP is robust.

\subsubsection{HazardEraPOP}
POP can be applied similarly to Hazard Eras.
Hazard Eras~\cite{ramalhete2017brief} has similar interface to Hazard Pointers, but unlike hazard pointers maintain a global monotonically increasing epoch variable. Each thread reserves and publishes the current value of this epoch when accessing a node, rather than reserving the node itself. Additionally, each node maintains its birth epoch and retire epoch, representing its lifespan during which it is reachable in a data structure. The key idea is that, before freeing a node, a thread compares the node’s lifespan to the reserved epochs. If no thread has reserved an epoch that intersects with the node’s birth and retire epochs, then no thread could hold a hazardous reference to the node, making it safe to free.
To obtain HazardEraPOP, we change the \Call{read}{} to only reserve epochs locally, and then publish for reclaimers by writing the reservations to shared memory when a reclaimer pings. 
The full pseudocode of HazardEraPOP along with its description and proof of safety and robustness appear in appendix \secref{apxhepop}.

\subsubsection{Note on a useful property of HazardPtrPOP}
HazardPtrPOP enables threads to privately track reservations using a lightweight \Call{read}{} and publish them on demand with a single fence when required by a \textit{reclaimer}. In essence, even if threads are stuck in a long-running execution, one can ping the stalled thread to learn which object it might currently be accessing. This feature allows us to develop an efficient variant of hazard pointers that approaches the performance of EBR. Specifically, in common cases, threads follow a fast path similar to EBR algorithms and when threads suspect delays, the publish-on-ping mechanism is used to assist a thread who is unable to reclaim.  
In the next section, we introduce this variant of hazard pointers that incorporates EBR.

% The ability of HazardPtrPOP to track reservations with lightweight \Call{read}{} (eliminating fence) and publish them on demand can be exploited to enhance the performance of HazardPtrPOP, bringing its performance closer to EBR while preserving the robustness of HazardPtrPOP.
% In the next section, we introduce a new algorithm that incorporates epochs and utilizes this property of HazardPtrPOP.

\subsection{EpochPOP}

\subsubsection{The overview}

In EpochPOP, threads operate in epochs, announcing their entry into and exit from the \textit{quiescent state} using a global epoch variable, similar to the EBR algorithm. However, unlike EBR, threads also privately track local reservations during traversals, utilizing the lightweight reads of HazardPtrPOP (no fence overhead), as if they are simulatenously executing HazardPtrPOP. This dual-mode operation occurs without the need for special synchronization (or alternation) between the two modes.

\textit{Reclaimers}, much like in EBR, periodically scan all announced epochs to free objects retired before the minimum announced epoch. In rare cases where thread delays prevent a \textit{reclaimer} from freeing its retired objects—such as when the retire list size remains above a certain threshold even after reclaiming in EBR mode—the \textit{reclaimer} employs publish-on-ping to force all threads to publish their current reservations. This action empties the retire list while skipping the reserved objects. This approach contrasts with techniques like neutralizing in NBR~\cite{singh2021nbr, singhTPDS2023NBRP} or ejecting in PEBR~\cite{kang2020marriage} and HP-BRCU~\cite{kim2024expediting}, in which signals are used to force threads to restart from programmer-defined checkpoints.

We describe EpochPOP as simultaneously operating in dual modes, deliberately avoiding the term \textit{fast-path slow-path} approach to avoid confusion with other techniques. Unlike fast-path slow-path techniques such as Qsense~\cite{balmau2016fast}, EpochPOP does not require switching between modes (and sychronizing between them). Instead, threads can seamlessly operate in both modes simultaneously, with different threads concurrently reclaiming objects in each mode.

\begin{algorithm}
\footnotesize
\caption{EpochPOP}
\label{algo:rcupophp}
    \begin{algorithmic}[1]
    % \Statex \textbf{global:}
    \State \texttt{const int reclaimFreq, epochFreq, C} \label{lin:rcuvar-epochfreq}
    \State \texttt{atomic<int> epoch} \label{lin:rcuvar-epoch}
    \State \texttt{thread\_local int tid, counter}
    \State \texttt{atomic<int> reservedEpoch[NTHREAD]} \label{lin:rcuvar-reservedepoch}
    \State \texttt{list<T*> retireList [NTHREAD]} \label{lin:rcuvar-retirelist}
        % \State \texttt{const epochFreq}
        % \State \texttt{const C}
    % \Statex
    % \Statex \textbf{per thread local structures:}
    \State \texttt{T* localReservations [NTHREAD][MAX\_HP]}
    \State \texttt{atomic<T*> sharedReservations [NTHREAD][MAX\_HP]} 
    \State \texttt{atomic<int> publishCounter [NTHREAD]}
    \State \texttt{int collectedPublishCounters [NTHREAD]}
        % \State \texttt{int counter} \Comment{to advance epoch}
    % \Statex
    % \Statex \textbf{per thread shared structures:}
    \Statex    
    \Procedure{startOp}{ }
        \If{\texttt{0 == ++counter \% epochFreq}}
            \State {\texttt{epoch.fetch\_add(1)}} \label{lin:incepoch}
        \EndIf
        \State {\texttt{reservedEpoch[tid] $\gets$ epoch}}
    \EndProcedure    
    
    \Statex
        % \Procedure{\texttt{T*} read}{\texttt{atomic<T*> \&ptrAddr, int slot}}
        %     \Repeat
        %         \State \texttt{T* ptr $\gets$ ptrAddr.load(memory\_order\_release)}
        %         \State \texttt{\textbf{localReservations[tid][slot] $\gets$ ptr}} \LComment{no store load fence needed.}
        %     \Until{\texttt{ptr $\ne$ ptrAddr.load(memory\_order\_release)}}
        % \EndProcedure
        \Procedure{\texttt{T*} read}{\texttt{atomic<T*> \&ptrAddr, int slot}} \label{lin:proc-read}
            \Repeat
                \State \texttt{T* readPtr $\gets$ *ptrAddr}
                \BeginBox[draw=black, dashed]
                \State \texttt{localReservations[tid][slot] $\gets$ readPtr}
                \EndBox
            \Until{\texttt{readPtr $=$ *ptrAddr}} \label{lin:loopexitep}
            \State \Return readPtr \label{lin:returnep}
        \EndProcedure    
    \Statex
        \Procedure{retire}{\texttt{T* ptr}}
            \State \texttt{myRetireList $\gets$ retireList[tid]}
            \State \texttt{myRetireList.append(ptr)}
            \State \texttt{ptr.retireEpoch $\gets$ epoch}
            % \State \texttt{listSize $\gets$ myRetireList.size()}

            \If {\texttt{$0 ==$ myRetireList.size() $\%$ reclaimFreq}}
                \State {\Call{reclaimEpochFreeable}{\texttt{myRetireList}}} \label{lin:reclaimepochfreeable}
                \If{\texttt{myRetireList.size() $\ge$ C*reclaimFreq}} \label{lin:pophpstylebegin}
                \BeginBox[draw=black, dashed]
                    \State {\Call{collectPublishedCounters}{ }} 
                    \State {\Call{\textbf{pingAllToPublish}}{ } } 
                    \State {\Call{waitForAllPublished}{ }}
                \EndBox
                % \State {\Call{collectAllReservations}{ }}
                \State {\Call{reclaimHPFreeable}{myRetireList}} \label{lin:rcureclaimhpfreeable}
                
                % \State \Call{pingAllToPublish}{ }
                % \State \Call{reclaimHPFreeable}{\texttt{myRetireList}}
                \EndIf
            \EndIf
        \EndProcedure    
        
    \Statex
        \Procedure{reclaimEpochFreeable}{\texttt{myRetireList}}
            \State { minReservedEpoch $\gets$ reservedEpoch.min()}
            \ForAll {\texttt{objPtr $\in$ myRetireList}}
                \If{\texttt{objPtr.retireEpoch} $<$ \texttt{minReservedEpoch}}
                    \State {\texttt{free(objPtr)}}
                \EndIf
            \EndFor
        \EndProcedure

    \Statex
        \Procedure{reclaimHPFreeable}{myRetireList}
                % \State \texttt{set<T*> collectedReservations $\gets$ \{\}}
                %     \ForAll {\texttt{<tid, slot> $\in$ sharedReservations[tid]}}
                %         \State {\texttt{objPtr $\gets$ sharedReservations[tid][slot]}}
                %         \State {\texttt{collectedReservations.insert(objPtr)}}
                %     \EndFor
                % \ForAll {\texttt{objPtr $\in$ myRetireList}}
                %     \If{\texttt{objPtr $\notin$ collectedReservations}}
                %         \State {\texttt{free(objPtr)}}
                %     \EndIf
                % \EndFor
                \LComment{Same as in \algoref{pophpcontd}}
        \EndProcedure
    \Statex
    \Procedure{endOp}{ }
        \State {\texttt{reservedEpoch[tid] $\gets$ MAX}}
        \State {\Call{clear}{ }} \label{lin:rcuclearall}
    \EndProcedure
    \end{algorithmic}
\end{algorithm}

\subsubsection{Description of algorithm}

\algoref{rcupophp} describes EpochPOP building upon an EBR implementation. All line reference in this section refer to \algoref{rcupophp}.
% ability to hide reservations during traversals building upon the implementation presented in the \algoref{rcu}.
Each thread maintains a \func{retireList} to collect objects retired by it (\lineref{rcuvar-retirelist}), a SWMR slot in \func{reservedEpoch} array to announce the current epoch it is executing in (\lineref{rcuvar-reservedepoch}), and a monotonically increasing \func{epoch} variable (\lineref{rcuvar-epoch}). 
The \func{epoch} is incremented periodically using the value of \func{epochFreq} (\lineref{rcuvar-epochfreq}).

Similar to original EBR, a thread announces it is exiting the quiescent state by reserving the current epoch at an appropriate slot in \func{reservedEpoch} with a call to \Call{startOp}{ }.
Additionally, to access new data structure objects within an operation, the thread uses \Call{read}{ }, privately reserving them (without fencing), similar to HazardPtrPOP.
Threads enter the \textit{quiescent state} by announcing the latest global epoch by invoking \Call{endOp}{ } and clearing the local reservations (\lineref{rcuclearall}).

\textit{Reclaimers} free nodes in their retire lists as they normally do in EBR (without considering threads' private reservations) unless a thread delay is suspected. During \Call{retire}{} calls, threads append retired objects to their retire lists by associating the current epoch as their \func{retireEpoch}. When the list reaches a threshold size, they invoke \Call{reclaimEpochFreeable}{} (\lineref{reclaimepochfreeable}). This procedure identifies the minimum epoch reserved by a thread from the \func{reservedEpoch} array and then frees the objects that were retired before that epoch.

If a \textit{reclaimer} suspects a thread is delayed, during the \Call{retire}{} call the reclaimer invokes the robust reclamation process of HazardPtrPOP.
That is, %More precisely, %in the example implementation,
if after attempting EBR style reclamation (\lineref{reclaimepochfreeable}), the reclaimer finds that its \func{retireList} is still too large (say, more than half of the \func{retireList} remains unreclaimed), %, i.e \texttt{C=1/2}),
it assumes that some thread reserving an older epoch has been delayed. The reclaimer therefore calls  \Call{reclaimHPFreeable}{} (\lineref{rcureclaimhpfreeable}) to
%This triggers reclamation in HazardPtrPOP style, where the \textit{reclaimer}
ping all threads, wait for reservations to be published, % to publish their reservations 
and free from its \func{retireList}. % using the \Call{reclaimHPFreeable}{} (\lineref{rcureclaimhpfreeable}).

% This is safe because EpochPOP, even though, in the common case, operates as EBR, privately tracks the objects it accesses. So, a \textit{reclaimer} when uses publish-on-ping can determine the objects which are safe to be freed.

Note that the private tracking of reservations is key in allowing reclaimers to continue to free objects in the presence of thread delays. % crucial  threads that are trying to reclaim objects, in the event where a delayed thread prevents freeing of it retired objects.
A thread that is delayed by other work will be interrupted by the reclaimer's signal, and will publish its reservations, allowing the reclaimer to precisely determine which objects the delayed thread will potentially access.
This allows the \textit{reclaimer} to safely free its retire list, skipping only a bounded set of reserved objects.

\subsubsection{Correctness and progress}

\begin{property}
    (Safety) EpochPOP avoids use-after-free errors. 
\end{property}
EpochPOP mostly runs classic EBR synchronization between \textit{readers} and \textit{reclaimers}.
In scenarios where no delayed threads are detected, \textit{reclaimers} only free objects whose retire timestamp indicates that they were retired before the oldest announced timestamp across all threads.
The success of the aforementioned condition indicates that all threads have gone quiescent at least once since the object (which a \textit{reclaimer} wishes to free) was retired. Consequently, no thread could hold a reference to this object.
In the event that a delayed thread is detected, a \textit{reclaimer} (1) signals all threads to timely publish the reservations they were maintaining all along (\asmref{sigasm}), (2) in a bounded waiting loop waits to establish that all reservation were published before it proceeds to free its retire list, (3) scan the reservations and frees only the objects that are not reserved (and thus cannot be accessed by any thread).
%This way, EpochPOP ensures no use-after-free errors occur.

% Note, while one thread could be reclaiming in popHP style, another thread could be reclaiming in epoch based reclamation style and the two threads concurrent reclaiming doesn't interfere with each other.

\begin{property}
    (Liveness) EpochPOP is robust.
\end{property}
This is ensured by the ability of the algorithm to maintain reservations while traversing the data structures and detect delayed threads. This allows threads to ensure continuous reclamation of objects in its retire list, only skipping a bounded set of reserved objects across all threads.

% . This enables popHP style reclamation, where a reclaimer could ping all threads to publish their reservations and subsequently free all retired nodes in its retire list  except at most $N*H$ nodes, which could be reserved by threads.

% \subsubsection{Note on applicability:}
% It can be noted that EpochPOP, like popHP, has to correctly reserve all the nodes before accessing. This involves reading a pointer to the node, reserving it locally, and then verifying that the reserved node was reachable at the time it was reserved. This implies that EpochPOP inherits the issue with applicability to data structures where logically deleted nodes are traversed~\cite{brown2015reclaiming}, like HP and popHP. However, ~\cite{jung2023applying} published a technique, called HP++, to address this issue of applicability in hazard pointers, and it will be worth exploring whether we can apply the same technique to popHP to extend its applicability.

% \begin{table*}
%     \centering
%     \begin{tabular}{cccc}
%     \toprule
%          & HP & RCU & popHP/popHPRCU\\
%     \midrule     
%         unreclaimed objects & bounded & unbounded & bounded\\
%         traversal progress & cond. lockfree & waitfree & cond. lockfree\\
%         reclamation progress & cond. lockfree & waitfree & cond. lockfree\\
%         traversal speed & low overhead & no or low overhead & low overhead\\        
%     \end{tabular}
%     \caption{Comparison of SMRs}
%     \label{tab:my_label}
% \end{table*}

\subsubsection{Ease of Use and Applicability}
The interface of the POP algorithms, including HazardPtrPOP, HazardEraPOP, and EpochPOP, is the same as that of hazard pointers. Additionally, these POP algorithms do not impose any extra usability limitations; thus, they are compatible with the same data structures as hazard pointers.
% Though we donot use the , Jung et. al.~\cite{jung2023applying} proposed a mechanism to extend the applicability of hazard pointers to data structures which permit traversal of logically deleted nodes like harris list~\cite{harris2001pragmatic}. 

% \textcolor{red}{Since POP does not introduce any additional usabil-
% ity constraints compared to hazard pointers, and QSBR can
% be applied to virtually any data structure, this means that
% QSense can be used with any data structure for which hazard
% pointers are applicable.}

% --\textcolor{blue}{We donot use Cadence and Dice et al's to compare because they assume or instrument nodes. bcse of there drawbacks. MENTION THIS IN EXP. Refer back to BG section}

\section{Experimental Evaluation}
\label{sec:experimentalEvaluation}

% \textcolor{red}{
We use the publicly available C++ benchmark from NBR(+)~\cite{singh2021nbr, singhTPDS2023NBRP}, which includes several safe memory reclamation algorithms: Hazard Pointers (\textbf{HP})~\cite{michael2004hazard}, Hazard Eras (\textbf{HE})~\cite{ramalhete2017brief}, \textbf{NBR+}~\cite{singh2021nbr, singhTPDS2023NBRP}, \textbf{IBR} and \textbf{EBR} (RCU-style as in IBR)~\cite{wen2018interval}. 
We extended the benchmark by implementing an optimized Linux \texttt{sys\_membarrier}-based version of HP (\textbf{HPAsym}, similar to Folly’s HP implementation) and adding our three algorithms: \textbf{HazardPtrPOP}, \textbf{HazardEraPOP}, and \textbf{EpochPOP}.
% and NBR~\cite{singhTPDS2023NBRP}. 
% although these results are not shown in the plots for the sake of clarity.
All algorithms were integrated with five data structures present in the benchmark: Brown’s (a,b)-tree (\textbf{ABT})~\cite{brown2017techniques}, the external binary search tree by David, Guerraoui, and Trigonakis (\textbf{DGT})~\cite{david2015asynchronized}, the Harris-Michael list (\textbf{HML})~\cite{michael2004hazard}, a lazy list (\textbf{LL})~\cite{heller2005lazy}, and a hashtable (\textbf{HMHT}) based on HML. These data structures account for several contention levels and memory access patterns. We also ran data structures without reclamation, which is shown as \textbf{NR} in the plots for a rough baseline.

The experiments for results presented in this paper were conducted on an Intel CascadeLake server - 144 threads, 4 NUMA nodes, 18 cores per node (2-way hyperthreaded), 99MiB L3 cache, 2.2 GHz frequency, 188GB RAM. 
Additional experiments are provided in the appendix (see \secref{extexp} and \secref{cyexp}), where we also compare with the recent Crystalline~\cite{nikolaev2024family} algorithm, showing that POP algorithms outperform it as well.
We also tested our benchmark on a 192 thread Intel Skylake SP, and obtained similar results. 

\subsubsection{Experimental Setting}

Our benchmark, compiled with \texttt{C++14} and \texttt{-O3} optimization, ran on Ubuntu 20.04 (kernel 5.8.0-55) using \texttt{numactl --interleave=all} and the mimalloc allocator. We used mimalloc (not jemalloc\cite{evans2006scalable}), as suggested by Brown et al.~\cite{kim2024token}, to avoid negative interaction between deferred memory reclamation and jemalloc, affecting scalability. Mimalloc\cite{leijen2019mimalloc}, with multilevel sharding for free lists, resolves this issue. Thus, we used mimalloc in our benchmark. Unless otherwise mentioned, we show results for DGT with a maximum size of 200K nodes, ABT with 20M size, lists (LL and HML) with 2K size and HMHT with 6M size and load factor of 6. We used 24K as max size of the retire list after which a reclamation event is triggered for all reclamation algorithms in all our experiments unless mentioned otherwise. \footnote{The NBR~\cite{singh2021nbr} authors used retire list of 32K to ensure fair comparison and avoid excessive signal overhead.}

\begin{figure*}
        \begin{subfigure}{\textwidth}
            \centering
            \resizebox{0.49\textwidth}{!}{\input{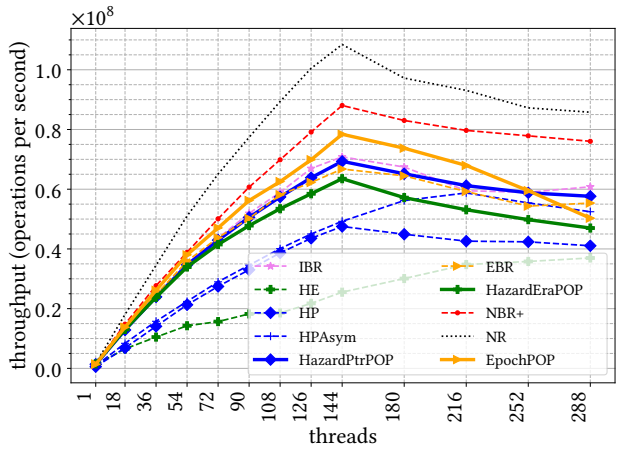}}
            \hfill
            \resizebox{0.49\textwidth}{!}{\input{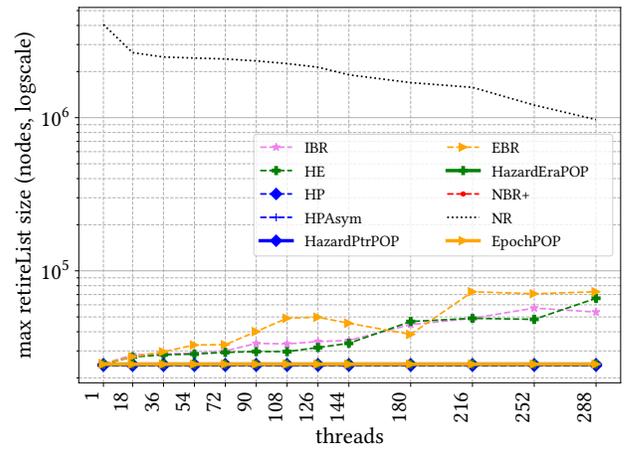}}
            \subcaption{Ext. Binary Search Tree (DGT). Size 200K}
            \label{fig:dgt100u}
        \end{subfigure}
        \hfill
        \begin{subfigure}{\textwidth}
            \centering
            \resizebox{0.49\textwidth}{!}{\input{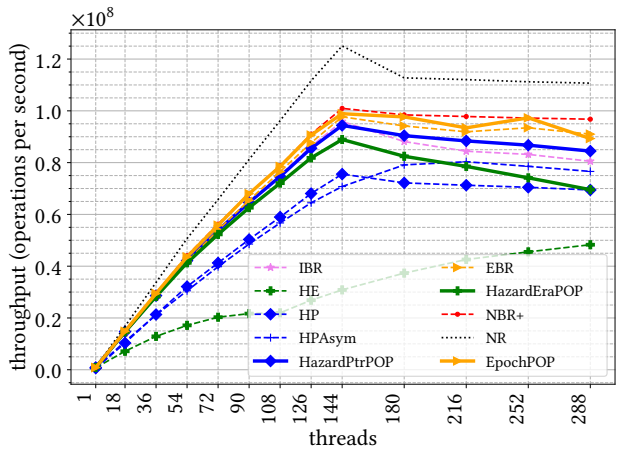}}
            \hfill
            \resizebox{0.49\textwidth}{!}{\input{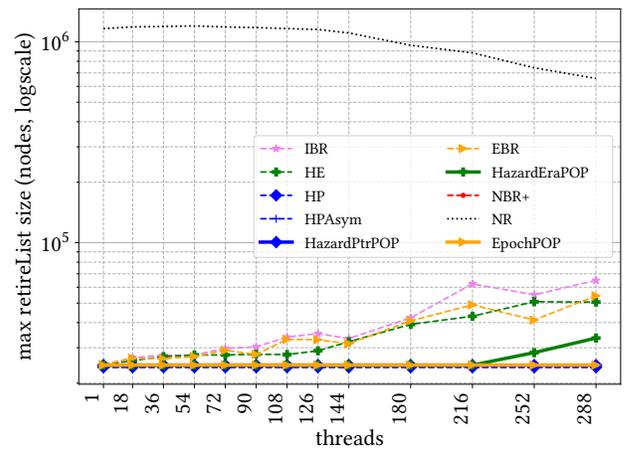}}
            \subcaption{Hash Table with Harris Michael List (HMHT). Size 6M.}
            \label{fig:hmht100u}
        \end{subfigure}
        \begin{subfigure}{\textwidth}
            \centering
            \resizebox{0.49\textwidth}{!}{\input{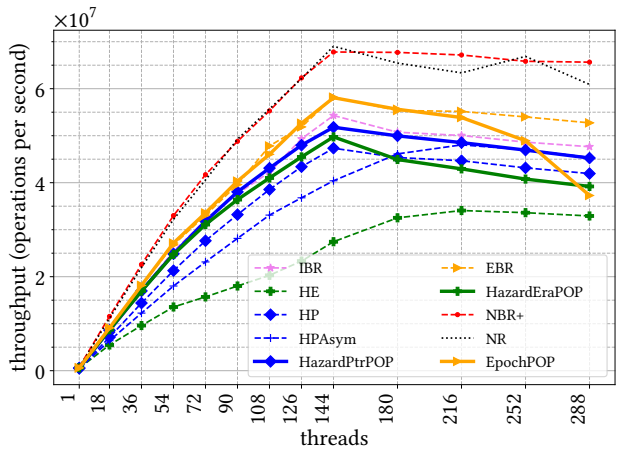}}
            \hfill
            \resizebox{0.49\textwidth}{!}{\input{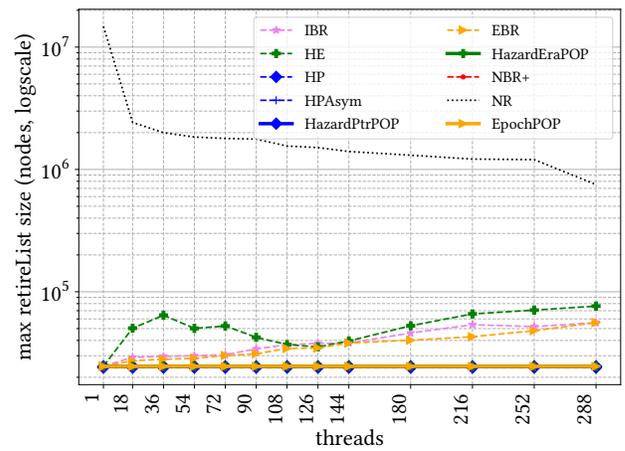}}
            \subcaption{(a,b) Tree (ABT). Size 20M.}
            \label{fig:abt100u}
        \end{subfigure}
        \caption{Workload: Update-heavy. Throughput and memory consumption across varying threads for different data structures.}
        \label{fig:exp1upd1}
\end{figure*}

\begin{figure*}
        \begin{subfigure}{\textwidth}
            \centering
            \resizebox{0.49\textwidth}{!}{\input{figs/hml/hml100u_2K_tput.pgf}}
            \hfill
            \resizebox{0.49\textwidth}{!}{\input{figs/hml/hml100u_2K_logy_maxbagsize.pgf}}
            \subcaption{Harris-Michael List (HML). Size 2K}
            \label{fig:hml100u}
        \end{subfigure}
        \begin{subfigure}{\textwidth}
            \centering
            \resizebox{0.49\textwidth}{!}{\input{figs/ll/ll100u_2K_tput.pgf}}
            \hfill
            \resizebox{0.49\textwidth}{!}{\input{figs/ll/ll100u_2K_logy_maxbagsize.pgf}}
            \subcaption{Lazy List(LL). Size 2K.}
            \label{fig:100u-ll}
        \end{subfigure}
        \caption{Workload: Update-heavy. Throughput and memory consumption across varying threads for different data structures.}
        \label{fig:exp1upd2}
\end{figure*}

\begin{figure*}
\centering
% \begin{minipage}{0.49\textwidth} % Set minipage width to 0.66 of the text width
%     \begin{subfigure}{\textwidth}
%         \centering
%         \resizebox{\linewidth}{!}{\input{figs/ll/ll100u_2K_tput.pgf}}\vfill
%         \caption{Throughput.}
%         \label{fig:100u-ll-tput}
%         \resizebox{\textwidth}{!}{\input{figs/ll/ll100u_2K_logy_maxbagsize.pgf}}
%         \caption{MemoryConsumption.}
%         \label{fig:100u-ll-mem}
%     \end{subfigure}
%     \caption{Workload: Update-heavy.\\Lazy List(LL). Size 2K.}
%     \label{fig:100u-ll}
% \end{minipage}
% \hfill
\begin{minipage}{0.49\textwidth} % Set minipage width to 0.66 of the text width
    \begin{subfigure}{\textwidth}
        \centering
        \resizebox{\linewidth}{!}{\input{figs/abt/abt10u_20M_tput.pgf}}\vfill
        \caption{ABT. Size 20M}
        \label{fig:10u-abt}
        \resizebox{\linewidth}{!}{\input{figs/dgt/dgt10u_200K_tput.pgf}}
        \caption{DGT. Size 200K}
        \label{fig:10u-dgt}
    \end{subfigure}
    \caption{Workload: Read-heavy. Throughput for ABT and DGT.}
    \label{fig:exp1read}    
\end{minipage}
\hfill
\begin{minipage}{0.49\textwidth}
    \begin{subfigure}{\textwidth}
        \centering
        \resizebox{\linewidth}{!}{\input{figs/lrr/hml50-longrun192jax_tput.pgf}}\vfill
        \caption{Throughput.}
        \label{fig:100u-hml-lrrtputratio}
        \resizebox{\linewidth}{!}{\input{figs/lrr/hml50-longrun192jax_logy_maxbagsize.pgf}}
        \caption{Memory Consumption.}
        \label{fig:100u-hml-lrrmem}
    \end{subfigure}
    \caption{Workload: Long Running Reads. HML list.}
    \label{fig:100u-hml-lrr}    
\end{minipage}
\end{figure*}

\subsubsection{Experimental Methodology}
In each trial of our experiment, threads prefill the data structure up to half of the maximum fixed key range (size). Subsequently, they enter an execution phase where, they perform data structure operations for 5 seconds. During execution, repeatedly a randomly chosen insert, delete, or contains operation is invoked with a key randomly selected from a given key range of the data structure.
We report throughput (millions of operations per second) and memory consumption (maximum garbage collected per thread) for read-heavy workloads with 90\% contains, 5\% inserts and 5\% deletes, as well as update-heavy workloads with 50\% inserts and 50\% deletes for thread in the range of 1 to 288. The system is oversubscribed after 144 threads.
The variance in the trials was below 5\%.
% Techniques using epochs increments it every $NTHREADS\times EPOCHFREQ$ allocation, where $EPOCHFREQ$ is a constant (set to 100 in our experiments).  

% We design our experiments with two objectives. First (EXP1, see \figref{e1}) is to evaluate the performance of our proposed publish-on-ping implementations HazardPOP, HazardErasPOP and EpochPOP. Second (EXP2, \figref{exp2}) aims to assess the peak memory consumption of publish-on-ping variants with and without stalled threads. For clarity, the publish-on-ping reclaimers are represented using solid lines in the plots.

\subsection{Results}
% \begin{figure*}
%     \centering
%     \includegraphics[width=0.49\linewidth, keepaspectratio]{figs/dgt50-jaxmem.png}\hfill
%     \includegraphics[width=0.49\linewidth, keepaspectratio]{figs/bge-dgt50mem.png}\hfill
%     % \includegraphics[width=0.49\linewidth, keepaspectratio]{figs/dgt50-jaxmem_logscale.png}\hfill
%     % \includegraphics[width=0.49\linewidth, keepaspectratio]{figs/bge-dgt50mem_logscale.png}\hfill
%     \caption{Exp2: Peak Memory Usage for DGT. Left: with no Stalled threads. Right: with stalled threads.}
%     \label{fig:exp2}
% \end{figure*}

In all the data structures across read-heavy and write-heavy workloads, publish-on-ping algorithms exhibit low memory footprint, and the impact of reduction in overhead due to the elimination of eager publishing of reservations with every pointer access translates to improvement in overall throughput and scalability.

\subsubsection{Write-heavy and Read-heavy Workload:}

For write-heavy workloads shown in \figref{exp1upd1} and \figref{exp1upd2}.
publish-on-ping (POP) algorithms consistently perform better or are similar and exhibit a lower memory footprint compared to the original algorithms on which they are based.
Specifically, HazardPtPOP is on average up to 70\% faster than HP and up to 20\% faster than HAasym. HazardEraPOP is on average up to 2x faster than HE. And
EpochPOP is similar in performance to EBR and IBR while maintaining strictly lower memory consumption. 
Note that at high thread counts, especially at 288 threads (oversubscribed), EpochPOP is slower in trees (DGT and ABT) but is comparable for in the hash table (HMHT). This predominantly is due to excessive signalling required to reclaim garbage, which is a reasonable trade-off to maintain a strictly lower memory footprint.

% \subsubsection{ReadHeavy Workload:}
In read-heavy workloads all POP algorithms are similar or, in some cases especially, at oversubscription marginally better than EBR and IBR, as shown in \figref{exp1read}. Individually, here again, HazardEraPOP is on average up to 3$\times$ faster than HE, HazardEpochPOP is marginally (up to ~15\%) better than HPAsym. 
\textit{Plots for additional data structures under read-heavy workloads can be found in the appendix \secref{extexp}.}

NBR+ demonstrates exceptional speed across various workloads and data structures due to the absence of read overhead and its minimal memory footprint, which improves cache performance. Although the POP algorithms eliminate the significant overhead of memory fences during reads, they still require the local reservation of pointers or eras, which are then published upon signaling for each read. 
This can be one reason for the slower performance of the POP algorithms compared to NBR+. 
However, as we will show in \secref{seclrr}, NBR+ incurs a high overhead, when retire list size is set very low (2K in our experiments), for long-running read operations (first shown by Kim, Jung and Kang in HP-BRCU~\cite{kim2024expediting}), leading to frequent restarts from entry points in data structures. This causes a substantial drop in read throughput. In contrast, our POP variants perform significantly faster in these scenarios, as they do not require threads to restart.
% \textcolor{red}{
Another notable advantage of HazardPtrPOP over NBR+ is its compatibility with some data structures~\cite{bronson2010practical, michael2004hazard} that cannot be paired with NBR+ but are compatible with HP~\cite{singhTPDS2023NBRP}, and hence with HazardPtrPOP. For these data structures, HazardPtrPOP can provide better performance than existing pointer reservation based algorithms.
% }
% Furthermore, in terms of compatibility, as mentioned in \ref{sec:} there are some data structures to which NBR+ doesnoy apply but our POP algorithms do.
% HPBRCU is implemented in Rust whihc makes a direct comparison

\subsubsection{Long Running Reads Workload:}
\label{sec:seclrr}

To demonstrate the advantage of POP algorithms over NBR+ for long-running reads, we conducted an experiment similar to the one in ~\cite{kim2024expediting} on a 192-thread Intel machine. We used the HML list with sizes ranging from 10K to 800K nodes. In this setup, 96 threads exclusively performed search operations, while another 96 threads executed update operations near the head of the list. We set the retire list size of all reclamation algorithms to 2K to frequently trigger reclamation events. This setup ensures that NBR+ frequently sends signals to restart the long-running read threads, resulting in a very low probability of completing reads.

\figref{100u-hml-lrr} presents the read throughput ratio (read throughput of other reclamation algorithms compared to NR) and memory consumption behavior for all algorithms in this experiment. The POP algorithms maintain high read throughput since they do not require read threads to restart when a thread reclaims memory, while also maintaining low memory consumption. In contrast, NBR+ slows down due to excessive restarts induced by reclaimers.

Subsequent techniques like HP-BRCU~\cite{kim2024expediting} and VBR~\cite{sheffi2021vbr} reduce the restart overhead for readers by introducing intermediate checkpoints, allowing threads to resume from the last checkpoint instead of the data structure’s entry point during long-running searches. However, HP-BRCU is implemented in Rust, and VBR requires a type-preserving memory allocator, making direct comparison difficult. Qualitatively, POP variants are superior as they avoid restarts entirely, while HP-BRCU and VBR still incur overhead, depending on checkpoint frequency.

% \cite{expeditionHP}
% -- Expediting HP showed signalling overhead causes throughput to crash.
% -- Due to signal overhead, the signal-based techniques are meant to be used with deferred reclamation and some tolerance with batch sie 16L to 32K. which is reasonable memory consumption.
% -- Issue with using long ops is that in our experiments for long running ops in lists with very large sizes the buffer never got filled o no reclamation occurred.
% -- To enable reclamation one needs to reduce the bagsize to get frequent signals which ofcourse kills perf.
% -- With our bag size and sig frequency we did not observe any perf issues. Max hML list for 200K size ran for 50 seconds and max bag size reported was $\sim$ 1K much less than bag size so no reclamation.. so no sig overhead... so retain perf. List tput at 20K size is really low.. and cannot retire upto bag size..

% -- So to see the massive consumption behavior, we use contrived workload.

% -- Show long running ops in appenidx and here the contrived workload.

% \subsubsection{Note on Long running OPerations}
% \subsubsection{Signal sensitivity Exp}
\section{Related Work}
\label{sec:relatedwork}
Research in solutions to the problem of safe memory reclamation for non-blocking data structures gained popularity in early 2000s with the appearance of epoch based~\cite{harris2001pragmatic, fraser2004practical, hart2007performance}, pointer reservation based techniques~\cite{michael2004hazard, herlihy2002repeat, herlihy2005nonblocking} and reference counting~\cite{valois1995lock, sundell2005wait, gidenstam2008efficient, detlefs2001lock, correia2021orcgc}. 
Since then, multiple manual and automatic techniques~\cite{cohen2015automatic, BlellochDRC21, anderson2022turning, cohen2018every, cohen2015efficient, sheffi2021vbr} have appeared with varying properties, such as performance, robustness, ease of use, and extent of applicability. Among these, some are designed with careful use of features in modern operating systems and architecture~\cite{jung2023applying, singh2021nbr, singh2023IPDPS, singhTPDS2023NBRP, dragojevic2011power, braginsky2013drop, kang2020marriage, singh2023efficient, alistarh2014stacktrack, alistarh2017forkscan, alistarh2018threadscan} while others leverage earlier techniques ~\cite{ramalhete2017brief, wen2018interval, anderson2021concurrent, nikolaev2019hyaline, nikolaev2020universal, nikolaev2021brief}. There are others which contribute to the improvement of earlier techniques~\cite{brown2015reclaiming, balmau2016fast, dice2016fast, moreno2023releasing}.
In this section, we exclusively focus on the deferred reservation-based techniques which aim to eliminate or reduce the overhead on the traversal path.

% Drop the Anchor (DTA)~\cite{braginsky2013drop} integrates epoch-based reclamation (EBR) and Hazard Pointers (HP). It primarily relies on EBR and enhances robustness against rare stalled threads by employing a recovery mechanism. This mechanism utilizes periodically published hazard pointers. During recovery, the process duplicates the range of objects reachable from the stalled thread into the data structure, enabling other threads to resume reclamation without having to worry about freeing objects reachable from the stalled threads.

One approach in~\cite{morrison2015temporally} assumes an alternative memory model called temporally bounded total store order (TBTSO) and applies it to HPs to guarantee that reserved pointers will be published to all threads within a bounded time, facilitating safe memory reclamation. 
In \cite{balmau2016fast}, Balmau et. al. employ context switches triggered by periodic scheduling of auxiliary processes per core to timely publish hazard pointers.
% In these techniques, reclaimers wait for a set interval of time to pass since an object is retired to ensure that reservations to the object (if any) would be published, reclaiming the object if it is not reserved. 
This technique incurs overhead to periodically publish reservations, even when threads might not be reclaiming.
Dice et. al.~\cite{dice2016fast} advocate using the write-protection feature of memory pages that triggers a global barrier to facilitate timely publication of hazard pointers before threads reclaim.
However, the issue with it is that it could block threads trying to reserve an object if a reclaimer stalls after write protecting the page the object resides on.   

Hazard Eras~\cite{ramalhete2017brief} reserves epochs which reduces the frequency of memory fences to publish reservations.
% , as reservations are published only when the global time stamp has changed since it was last reserved.
Although the frequency of memory fences is reduced, the overhead remains substantial as seen in our experiments. Another technique, Conditional Access~\cite{singh2023IPDPS} utilizes hardware-software co-design to eliminate explicit synchronization required between reclaimers and readers at the programming level by leveraging existing synchronization at the cache coherence level. However, this might not be available in hardware any time soon. 
% This allows conditional access to be fast without delaying reclamation at all.

The question of determining which objects are safe to free compelled earlier techniques using signals, such as DEBRA+~\cite{brown2015reclaiming} and NBR~\cite{singh2021nbr} to forcibly change control flow, compelling threads to restart, inducing overheads for long-running read operations such as in OLTP workloads~\cite{kim2024expediting}. However, EpochPOP \textit{does not} need to alter the control flow of threads, due to its ability to track reservations privately and publish them on demand.

Kim, Jung and Kang proposed HP-RCU and HP-BRCU~\cite{kim2024expediting} that execute in a sequence of HP and RCU phases and amortize the per read traversal cost by HP-based checkpoints from which traversals have to restart in case signalled for reclamation. Compared to NBR~\cite{singh2021nbr}, HP-BRCU turns the coarse-grained restart overhead into fine-grained with the use of intermediate checkpoints. But, like other checkpoint-based techniques (VBR\cite{sheffi2021vbr}), it could be complex to use and requires programmers to identify regions in the code for safe application.
Furthermore, installing checkpoints determining safe points in data structure code to enable restart may be difficult for arbitrary data structures~\cite{Cohen_thesis16}.
POP algorithms on the other hand are simpler, are able to avoid per-node overhead without inducing signal-based rollbacks and retain the original programmability of hazard pointers.

\section{Conclusion}
\label{sec:conclusion}
% \textcolor{blue}{
We have proposed the publish-on-ping approach for reclamation, utilizing POSIX signals to expedite pointer based techniques like hazard pointers and hazard eras. 
It can be used easily as a replacement for hazard pointers\textendash a technique set to be included in the C++26 standard library. Furthermore, we integrate epochs alongside the publish-on-ping variant of hazard pointers. The resulting algorithm, EpochPOP, performs similar to epoch-based reclamation, while providing bounded garbage.
Overall, publish-on-ping algorithms retain ease of programming in comparison to prior signal-based approaches, like DEBRA+ and NBR, as the former do not necessitate a reclamation-triggered change in control flow in data structure operations.
In the future, it would be worth exploring whether other synchronization problems exhibiting a safe memory reclamation-like pattern can leverage our approach.
\begin{acks}
Supported by the Hellenic Foundation for Research and Innovation (HFRI) under the ``Second Call for HFRI Research Projects to support Faculty Members and Researchers'' (project number: 3684).
This work was also supported by the Natural Sciences and Engineering Research Council of Canada (NSERC) Collaborative
Research and Development grant: 539431-19, the Canada
Foundation for Innovation John R. Evans Leaders Fund (38512)
with equal support from the Ontario Research Fund CFI Leaders Opportunity Fund, NSERC Discovery Program Grant:
2019-04227, NSERC Discovery Launch Grant: 2019-00048,
and the University of Waterloo. The findings and opinions
expressed in this paper are those of the authors and do not
necessarily reflect the views of the funding agencies. We
also thank the anonymous reviewers for their thoughtful
feedback. 
\end{acks}

%%
%% The next two lines define the bibliography style to be used, and
%% the bibliography file.
\bibliographystyle{ACM-Reference-Format}
\bibliography{sample-base}

%%
%% If your work has an appendix, this is the place to put it.
% \newpage
\appendix
\section{Artifact Description}
\label{sec:aedescp}
This section provides a step by step guide to run our artifact in a docker container.

The artifact can be found at the following links:
\begin{itemize}
    \item zenodo (with docker image): \\
    \url{https://doi.org/10.5281/zenodo.14219155}.
    \item gitlab (repo without docker image):\\
    \url{https://gitlab.com/aajayssingh/pop_setbench}
\end{itemize}

If you prefer to use the artifact directly without using the docker container please refer to the accompanying README file in the source code.

% To better reproduce the results described in our paper we recommend to run the pop\_setbench on a NUMA machine with at least 2 NUMA nodes having a recent Linux distro (we used Ubuntu 18.04 or 20.04) with 188GB RAM and recent Docker installation (we used version 19.03.6, build 369ce74a3c). 

The following instructions will help you load and run the provided Docker image within the artifact downloaded from Zenodo link.
Once the docker container starts you can use the accompanying README file to compile and run the experiments in the benchmark.
\\
\\
\myparagraph{Steps to load and run the provided Docker image:}

Note: Sudo permission may be required to execute the following instructions. 

\begin{enumerate}
    \item 
    Install the latest version of Docker on your system. We tested the artifact with the Docker version 24.0.7, build 24.0.7-0ubuntu2~20.04.1. Instructions to install Docker may be found at \url{https://docs.docker.com/engine/install/ubuntu/}. Or you may refer to the “Installing Docker” section at the end of this README.

    To check the version of docker on your machine use:
\begin{verbatim}
$ docker -v
\end{verbatim}

    \item Download the artifact from Zenodo at URL:\\
    \url{https://doi.org/10.5281/zenodo.14219155}.
    % Alternatively, most recent version could be downloaded at the gitlab repo, URL: \url{https://gitlab.com/aajayssingh/nbr_setbench}

    \item Extract the downloaded folder and move to \\
    \textit{pop\_setbench/} directory using $cd$ command.
    \item Find docker image named \textit{pop\_docker.tar.gz}\\
    in pop\_setbench/ directory. And load the downloaded docker image with the following command:
\begin{verbatim}
 $ sudo docker load -i pop_docker.tar.gz
\end{verbatim}
\item 
Verify that image was loaded:
\begin{verbatim}
 $ sudo docker images
\end{verbatim}
\item Start a docker container from the loaded image:
\begin{verbatim}
 $ sudo docker run --name pop -i -t \ 
 --privileged pop_setbench /bin/bash
\end{verbatim}
\item Invoke $ls$ to see several files/folders of the artifact: Dockerfile, README.md, common, ds, install.sh, lib, microbench, pop\_experiments, tools.
\end{enumerate}

Now, to compile and run the experiments you could follow the instructions in the README file.

\newpage
%#########################
\section{Hazard Eras and HazardEraPOP}
%#########################

\subsection{Hazard Eras}
\label{sec:apxhe}

\begin{algorithm}[h]
    \caption{Original Hazard Era ~\cite{ramalhete2017brief, wen2018interval}.}
    \label{algo:orighe}
    \begin{algorithmic}[1]
    \State \texttt{const reclaimFreq} \Comment{frequency of reclaiming retire list} 
    \State \texttt{thread\_local int tid} \Comment{current thread id}
    \State \texttt{list<T*> retireList [NTHREAD]} 
        \State \texttt{atomic<int> sharedReservations [NTHREAD][MAX\_HE]} 
        \State \texttt{atomic<int> epoch} \Comment{incremented periodically}
        \State \texttt{int collectedReservations[NTHREAD][MAX\_HE]$\gets$\{\}} 
        \Statex
        \Procedure{\texttt{T*} read}{\texttt{atomic<T*> \&ptrAddr, int slot}} \label{lin:orige-procread}
            \State \texttt{int oldEra $\gets$ sharedReservations[tid][slot]} 
            \While{\textit{True}}
                \State \texttt{T* readPtr $\gets$ *ptrAddr}
                \State \texttt{int newEra $\gets$ epoch}                
                \If{oldEra == newEra}
                    \Return readPtr
                \EndIf
            \State \texttt{sharedReservations[tid][slot] $\gets$ newEra} \LComment{store load fence.}
            \State \texttt{oldEra $\gets$ newEra} 
            \EndWhile
        \EndProcedure
    \Statex
        \Procedure{retire}{\texttt{T* ptr}} \label{lin:proc-origheretire}
            \State \texttt{myRetireList $\gets$ retireList[tid]}
            \State \texttt{myRetireList.append(ptr)}
            \State \texttt{ptr.retireEpoch $\gets$ epoch}
            \If {\texttt{myRetireList.size() $\geq$ reclaimFreq}}
                \State \texttt{epoch.fetch\_and\_add(1)}
                % \State {\Call{collectAllReservations}{ }}
                \State {\Call{reclaimHEFreeable}{myRetireList}}\label{lin:rechefreeable}
            \EndIf
        \EndProcedure    
    \Statex
        \Procedure{clear}{ } \label{lin:orighe-proc-clearall}
            \For{$slot=0, \dots, MAX\_HE$}
                    \State \texttt{localReservations[tid][slot] $\gets$ NONE}
                    % \LComment{store load fence}
            \EndFor        
        \EndProcedure        

% \algstore{algpophe}
% \end{algorithmic}
% \end{algorithm}

% \begin{algorithm} [H]                    
% \begin{algorithmic} [1]                   % enter the algorithmic environment
% \algrestore{algpophe}
    \Statex
        \Procedure{reclaimHEFreeable}{myRetireList} 
                \LComment{collect all published reservations}
                % \ForAll {\texttt{tid}}
                    \ForAll {\texttt{<tid, slot> $\in$ sharedReservations[tid]}} \label{lin:orighecollectstart}
                        \State {\texttt{reservedEra $\gets$ sharedReservations[tid][slot]}}
                        \State {\texttt{collectedReservations[tid][slot] $\gets$ reservedEra}}
                    \EndFor \label{lin:orighecollectend}
                % \EndFor                
                \LComment{free all nodes not reserved}
                \ForAll {\texttt{objPtr $\in$ myRetireList}} \label{lin:orighereclaimstart}
                    \If{\Call{canFree}{objPtr}} \label{lin:orighecanfree}
                        \State {\texttt{free(objPtr)}}
                    \EndIf
                \EndFor \label{lin:orighereclaimend}
        \EndProcedure
        \Statex
        \Procedure{canFree}{objPtr}
            \ForAll {\texttt{<tid, slot> $\in$ collectedReservations[tid]}}
                \State {\texttt{reservedEra $\gets$ collectedReservations[tid][slot]}}
                \If{reservedEra $<$ objPtr.birthEra OR reservedEra $>$ objPtr.retireEra OR reservedEra == NONE}\label{lin:orighecanfreecomp}
                    \State \textit{continue}
                    % \Statex
                \EndIf
                \Return{False}
            % \Statex
            \EndFor
            \Return{True}
        \EndProcedure
\end{algorithmic}
\end{algorithm}

\subsection{HazardEraPOP}
\label{sec:apxhepop}

We apply pubish-on-ping to HE that results in the algorithm called HazardEraPOP. 

\begin{algorithm}
    \caption{HazardEraPOP: Hazard Era Publish-on-Ping.}
    \label{algo:pophe}
    \begin{algorithmic}[1]
    \State \texttt{const reclaimFreq} \label{lin:hevar-reclaimfreq}\Comment{frequency of reclaiming retire list} 
    \State \texttt{thread\_local int tid} \Comment{current thread id}
    \State \texttt{list<T*> retireList [NTHREAD]} \label{lin:hevar-retirelist}
        \State \texttt{int localReservations [NTHREAD][MAX\_HE]} 
        \State \texttt{atomic<int> sharedReservations [NTHREAD][MAX\_HE]} \label{lin:hevar-sharedreservations}
        \State \texttt{atomic<int> publishCounter [NTHREAD]} \label{lin:hevar-publishcounter}
        \State \texttt{thread\_local int collectedPublishCounters [NTHREAD]}
        \State \texttt{int collectedReservations[NTHREAD][MAX\_HE]$\gets$\{\}} 

        \Procedure{\texttt{T*} read}{\texttt{atomic<T*> \&ptrAddr, int slot}}
            \State \texttt{int oldEra $\gets$ localReservations[tid][slot]} 
            \While{\textit{True}}
                \State \texttt{T* readPtr $\gets$ *ptrAddr}
                \State \texttt{int newEra $\gets$ epoch}                
                \If{oldEra == newEra}
                    \Return readPtr \label{lin:hevalsucc}
                \EndIf
            \State \texttt{localReservations[tid][slot] $\gets$ newEra} \label{lin:lochereserve}\LComment{no store load fence needed.}
            \State \texttt{oldEra $\gets$ newEra} 
            \EndWhile
        \EndProcedure
    \Statex
        \Procedure{retire}{\texttt{T* ptr}} \label{lin:heproc-retire}
            \State \texttt{myRetireList $\gets$ retireList[tid]}
            \State \texttt{myRetireList.append(ptr)}
            \State \texttt{ptr.retireEpoch $\gets$ epoch}
            \If {\texttt{myRetireList.size() $\geq$ reclaimFreq}}
                \BeginBox[draw=black, dashed]
                    \State {\Call{collectPublishedCounters}{ }}\label{lin:hecollectpublishedcounters}
                    \State {\Call{\textbf{pingAllToPublish}}{ } } \label{lin:hepingalltopublish}
                    \State {\Call{waitForAllPublished}{ }} \label{lin:hewaitforallpublished}
                \EndBox
                % \State {\Call{collectAllReservations}{ }}
                \State {\Call{reclaimHEFreeable}{myRetireList}}
            \EndIf
        \EndProcedure    
    \Statex
        \Procedure{clear}{ } \label{lin:he-proc-clearall}
            \For{$slot=0, \dots, MAX\_HE$}
                    \State \texttt{localReservations[tid][slot] $\gets$ NONE} \label{lin:setnone}
                    % \LComment{no store load fence needed}
            \EndFor        
        \EndProcedure 
        \Statex
        \Procedure{reclaimHEFreeable}{myRetireList} \label{lin:heproc-reclaimhefreeable}
                \LComment{collect all published reservations}
                % \ForAll {\texttt{tid}}
                    \ForAll {\texttt{<tid, slot> $\in$ sharedReservations[tid]}}\label{lin:he-freestart}
                        \State {\texttt{reservedEra $\gets$ sharedReservations[tid][slot]}}
                        \State {\texttt{collectedReservations[tid][slot] $\gets$ reservedEra}}
                    \EndFor \label{lin:he-collectend}
                % \EndFor                
                \LComment{free all nodes not reserved}
                \ForAll {\texttt{objPtr $\in$ myRetireList}} 
                    \If{\Call{canFree}{objPtr}}
                        \State {\texttt{free(objPtr)}}
                    \EndIf
                \EndFor \label{lin:he-freeend}
                % \LComment{Same as in \algoref{orighe}}
        \EndProcedure
    \Statex
        \Procedure{canFree}{objPtr}
            \ForAll {\texttt{<tid, slot> $\in$ collectedReservations[tid]}}
                \State {\texttt{reservedEra $\gets$ collectedReservations[tid][slot]}}
                \If{reservedEra $<$ objPtr.birthEra OR reservedEra $>$ objPtr.retireEra OR reservedEra == NONE}
                    \State \textit{continue}
                \EndIf
                \State \textit{return False}
            \EndFor
            \State \textit{return True}
            % \LComment{Same as in \algoref{orighe}}
        \EndProcedure
% \algstore{algpophe}
\end{algorithmic}
\end{algorithm}

\algoref{pophe} shows the implementation of the proposed HazardEraPOP algorithm. From a programmer's perspective, it maintains the same interface as HP, HE, or HazardPointersPOP. However, unlike the eager publishing reservations paradigm in HE, threads in HazardEraPOP save the nodes (to reserve) locally in their \func{localReservations} array during \Call{read}{}.

The \Call{read}{} procedure fetches the pointer to the node currently being read, reads the value of the current epoch, and compares it with the previously saved epoch value in the slot at \func{localReservations}. 
The slot corresponds to the current pointer being fetched.
If the previously locally reserved epoch and matches the current epoch, i.e. the current epoch has not changed while the pointer to the node was being read, then the node is safe to be read, and the pointer to the node is returned. 
Otherwise, the new epoch value is locally reserved (in \func{localReservations} without publishing, at \lineref{lochereserve}) and the process is repeated until the global epoch remains unchanged while the value at the input atomic pointer containing the address of the node currently being read is successfully fetched (\lineref{hevalsucc}).

When a thread is about to go \textit{quiescent}, i.e., while exiting the data structure operation, it resets the local reservations by setting the corresponding slots to NONE (\lineref{setnone}) using \Call{clear}{}.
These local reservations are then published to a shared array of single-writer multi-reader slots, called \func{sharedReservations} (\lineref{hevar-sharedreservations}), when pinged by a reclaimer.

Threads maintain a per thread list, depicted as \func{retireList} (\lineref{hevar-retirelist}), to which they append the retired nodes with call to \Call{retire}{ } during their update operations.
When the list size exceeds a threshold set in \func{reclaimFreq} (\lineref{hevar-reclaimfreq}), a \textit{reclaimer} invokes \Call{pingAllToPublish}{ } to trigger the publishing of local reservations. This function is same as in original HE algorithms~\algoref{orighe}.

% In \Call{pingAllToPublish}{ } (\lineref{heproc-pingalltopublish}) a \textit{reclaimer} uses \func{pthread\_kill()} to ping all threads.
% The threads receiving these pings execute a signal handler, called \Call{publishReservations}{ } (\lineref{heproc-publishreservations}), where they write all their local reservations to the shared array.
% Once all threads have completed execution of \Call{publishReservations}{}, the reservations become visible to the \textit{reclaimer}.

After every thread publishes its reservations, the \textit{reclaimer} can invoke \Call{reclaimHEFreeable}{ } to free all retired nodes that are not reserved (\lineref{heproc-reclaimhefreeable}). Specifically, within \Call{reclaimHEFreeable}{ }, the \textit{reclaimer} collects all reservations and frees those that are not in the collected reservations.  This function is same as in original HE algorithms~\algoref{orighe}.

The functions \Call{publishReservations}{ }, \Call{collectPublishedCounters}{ }, 
\Call{pingAllToPublish}{ }, and \Call{waitForAllToPublished}{ } are the same as shown in the HazardPtrPOP algorithm.

\subsubsection{Correctness and Progress in HazardEraPOP}

\begin{property}
    (Safety) HazardEraPOP is safe from use-after-free errors. 
\end{property}
In order to prove that HazardEraPOP is safe, we need to establish that any \textit{reclaimer} will not free a node that other threads could subsequently access.
$Wlog.$, by the way of contradiction, let us assume that a thread $T1$ frees a node $n$ at a time $t1$ which is subsequently accessed by another thread $T2$ at a later time $t2$ ($t1 < t2$). 

In order to access $n$, $T2$ must successfully protect it at an earlier time $t2'$, such that $t2' < t2$. 
Similarly, to free $n$, $T1$ retires, then pings all threads to publish their reservations, and then waits, for a bounded time, to ensue that all threads complete publishing their reservations at a time $t1'$, such that $t1' < t1$.

Now, two cases arise. 
First, $t1' < t2'$, that is, $T2$ published all its reservations for $T1$ to see before $T2$ reserved $n$ at $t2'$. This is only possible if $T2$ reserved the retired $n$, which is not possible because $T2$ will fail to reserve $n$ at $t2'$ because $T1$ before pinging would have changed the global epoch. 
In the second case, $t2' < t1'$, that is, $n$ was successfully reserved before $T1$ requested $T2$ to publish the reservations.
Note, in this case, $T2$ would have published the reservation of the epoch value at the time the pointer to $n$ is read, and $T1$ will scan the reservation and is guaranteed to find the reserved epoch. Now, when $T1$ attempts to free $n$ it will notice that the reserved epoch is less or equal to the retire epoch, i.e. lifespan of $n$ overlaps with the reserved epoch and there will not free $n$.
Hence, HazardEraPOP is immune to use-after-free errors.

\begin{property}
    (Liveness) HazardEraPOP is robust.
\end{property}
HazardEraPOP retains the original robustness property of Hazard Eras. A reserved epoch could only prevent reclamation of the nodes whose lifespan intersects with the epoch. All nodes allocated after or retired before the reserved epoch can continue to be reclaimed.

\section{RCU style implementation of Epoch Based Reclamation}
\label{sec:apxrcu}

\begin{algorithm}
\caption{Epoch Based Reclamation~\cite{wen2018interval} upon which EpochPOP algorithm is based.}
\label{algo:rcuebr}
    \begin{algorithmic}[1]
    % \Statex \textbf{global:}
        \State \texttt{const reclaimFreq, epochFreq} \label{lin:var-reclaimFreq} \label{lin:var-epochFreq} 
        \State \texttt{atomic<int> epoch} \Comment{incremented periodically} \label{lin:epoch}
        % \State \texttt{const epochFreq} 
    % \Statex
    % \Statex \textbf{per thread local:}
        \State \texttt{thread\_local int counter, tid} \Comment{tid is current thread id}
        % \State \texttt{int } \Comment{to advance epoch}
    % \Statex
    % \Statex \textbf{per thread shared:}
    \State \texttt{atomic<int> reservedEpoch[NTHREAD]} \label{lin:var-reservedEpoch}
    \State \texttt{list<T*> retireList [NTHREAD]} \label{lin:var-retireList}

    \Statex    
    \Procedure{startOp}{ } \label{lin:proc-startOp}
        \If{\texttt{0 == ++counter \% epochFreq}}
            \State {\texttt{epoch.fetch\_add(1)}} \label{lin:incrementEpoch}
        \EndIf
        \State {\texttt{reservedEpoch[tid] $\gets$ epoch}} \label{lin:reservEpoch}
    \EndProcedure        

    \Statex
        \Procedure{retire}{\texttt{T* ptr}}
            \State \texttt{myRetireList $\gets$ retireList[tid]}
            \State \texttt{myRetireList.append(ptr)}
            \State \texttt{ptr.retireEpoch $\gets$ epoch}
            \If {\texttt{$0 ==$ myRetireList.size() $\%$ reclaimFreq}}
                \State {\Call{reclaimEpochFreeable}{\texttt{myRetireList}}}
            \EndIf
        \EndProcedure    
        
    \Statex
        \Procedure{reclaimEpochFreeable}{\texttt{myRetireList}} \label{lin:proc-recEpochFree}
            \State { minReservedEpoch $\gets$ reservedEpoch.min()}
            \ForAll {\texttt{objPtr $\in$ myRetireList}}
                \If{\texttt{objPtr.retireEpoch} $<$ \texttt{minReservedEpoch}}
                    \State {\texttt{free(objPtr)}}
                \EndIf
            \EndFor
        \EndProcedure

    \Statex
    \Procedure{endOp}{ }
        \State {\texttt{reservedEpoch[tid] $\gets$ MAX}}
    \EndProcedure
    \end{algorithmic}
\end{algorithm}

\onecolumn
\section{Experiments on 144 thread Intel CascadeLake Server.}
\label{sec:extexp}

From Intel’s CascadeLake server --- 144 threads, 4 NUMA nodes,
18 cores per node (2-way hyperthreaded), 99MiB L3 cache, 2.2 GHz
frequency, 188GB RAM. 

\begin{figure*}[h]
\centering
     \begin{minipage}{\textwidth}
        \begin{subfigure}{\textwidth}
            \includegraphics[width=0.33\linewidth, keepaspectratio]{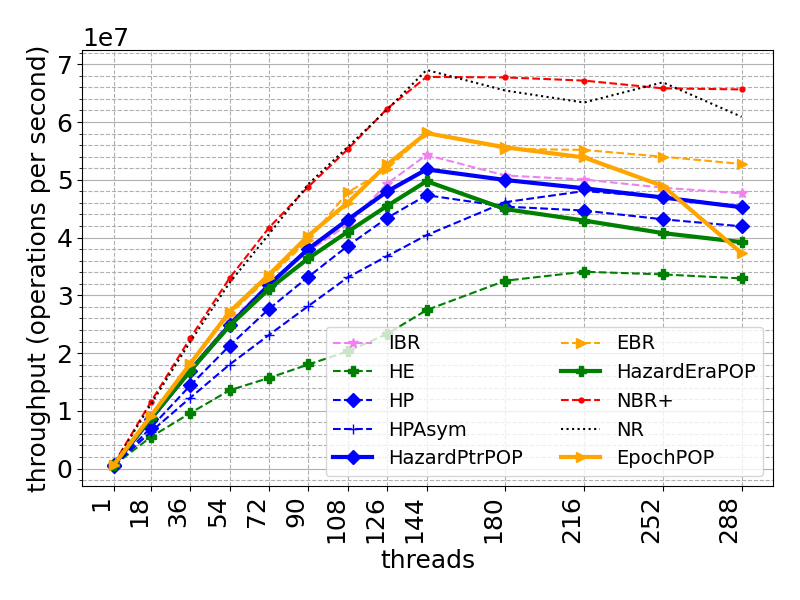}\hfill
            \includegraphics[width=0.33\linewidth, keepaspectratio]{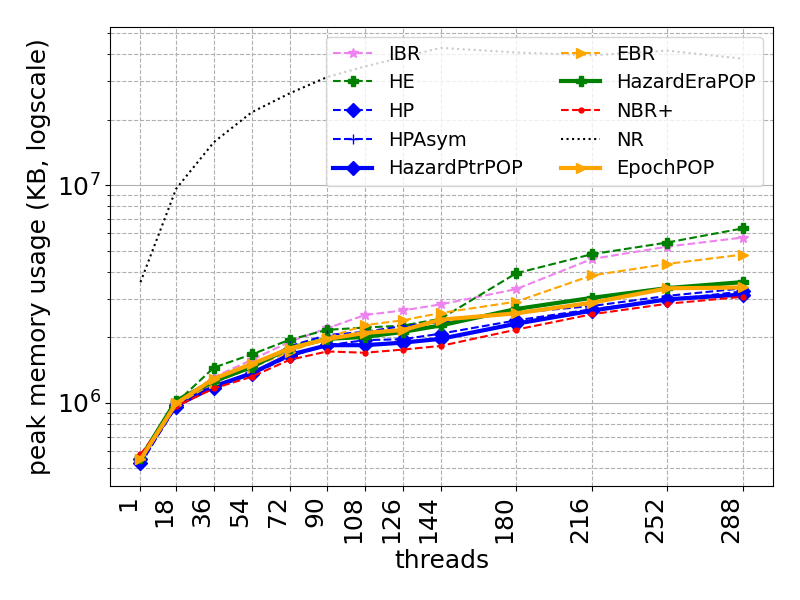}\hfill
            \includegraphics[width=0.33\linewidth, keepaspectratio]{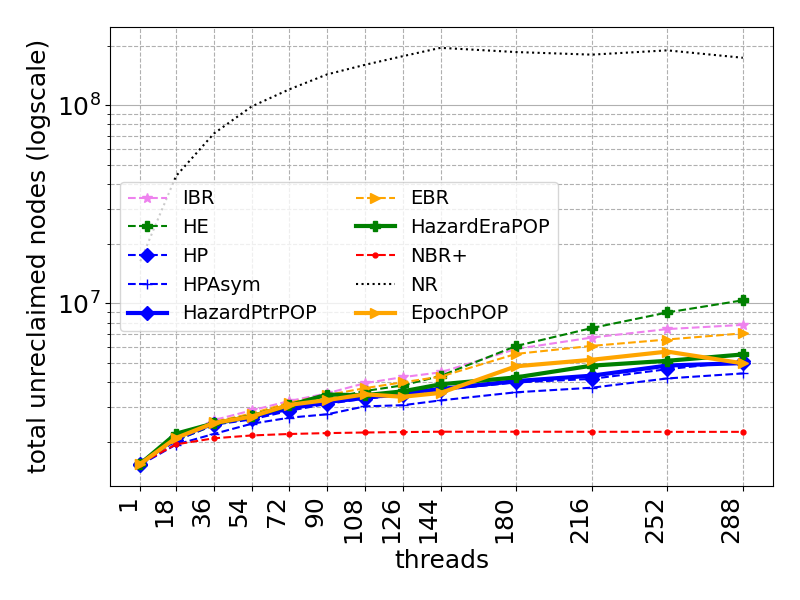}\hfill
            \caption{update heavy: 50\% inserts and 50\% deletes. }
            \label{fig:abt100u}
        \end{subfigure}
        \begin{subfigure}{\textwidth}
            \includegraphics[width=0.33\linewidth, keepaspectratio]{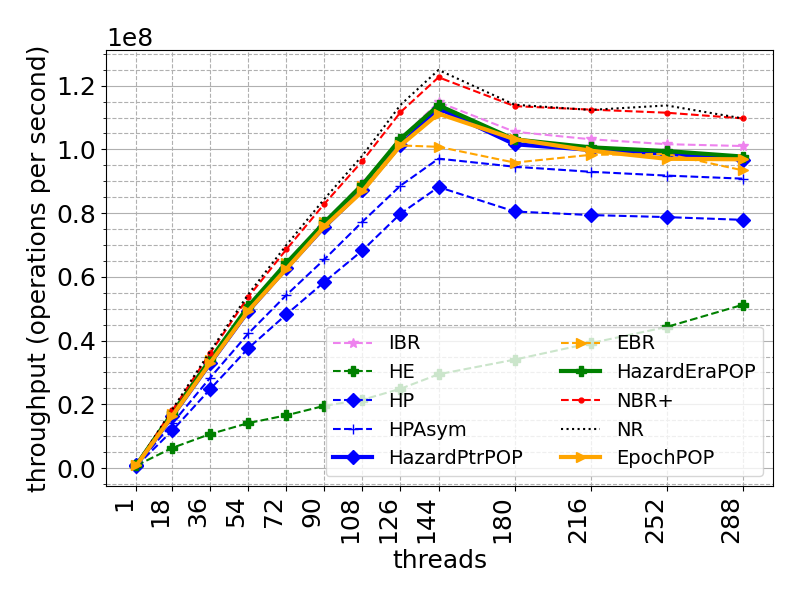}\hfill
            \includegraphics[width=0.33\linewidth, keepaspectratio]{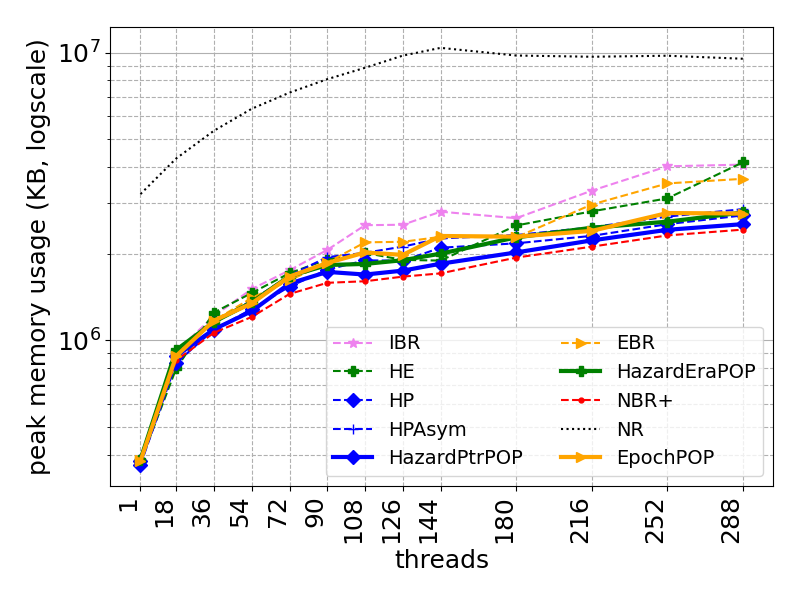}\hfill
            \includegraphics[width=0.33\linewidth, keepaspectratio]{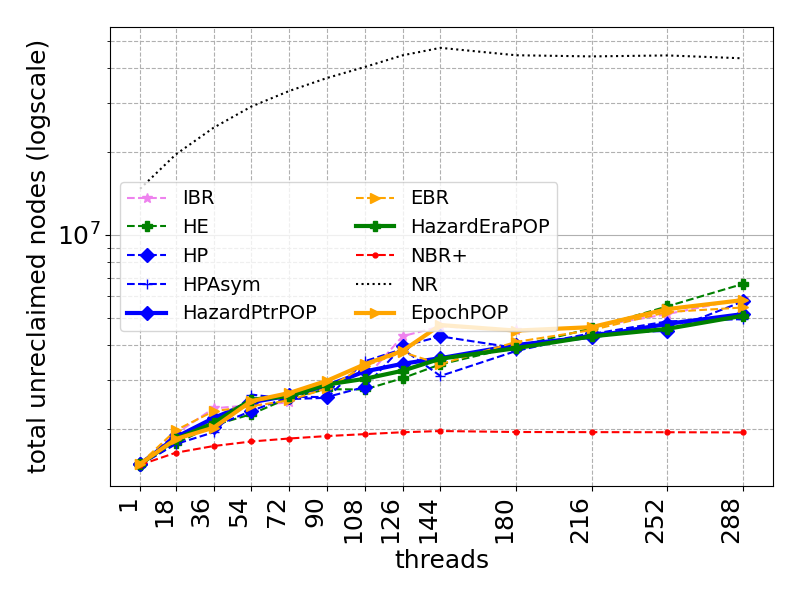}\hfill
            \caption{read heavy: 5\% inserts, 5\% deletes and 90\% contains.}
            \label{fig:abt10u}
        \end{subfigure}
     \end{minipage}
    \label{fig:abt}
    \caption{ABT. Size 20M. [Left: Throughput]. [Center: max resident memory in system]. [Right:total unreclaimed nodes].}
\end{figure*}

\begin{figure*}[h]
\centering
     \begin{minipage}{\textwidth}
        \begin{subfigure}{\textwidth}
            \includegraphics[width=0.33\linewidth, keepaspectratio]{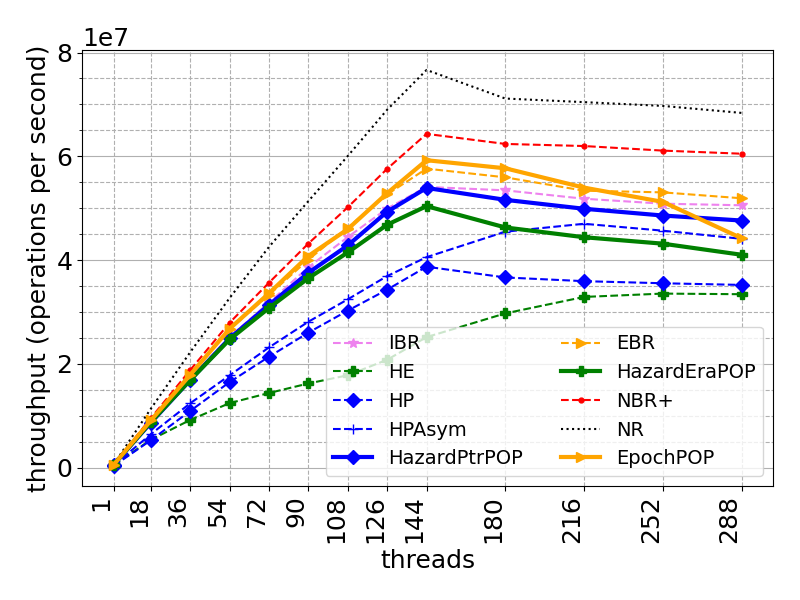}\hfill
            \includegraphics[width=0.33\linewidth, keepaspectratio]{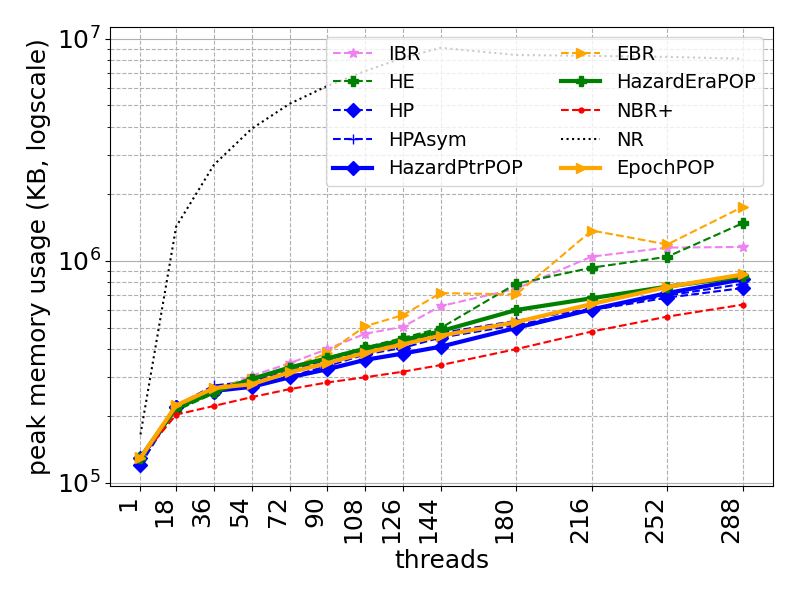}\hfill
            \includegraphics[width=0.33\linewidth, keepaspectratio]{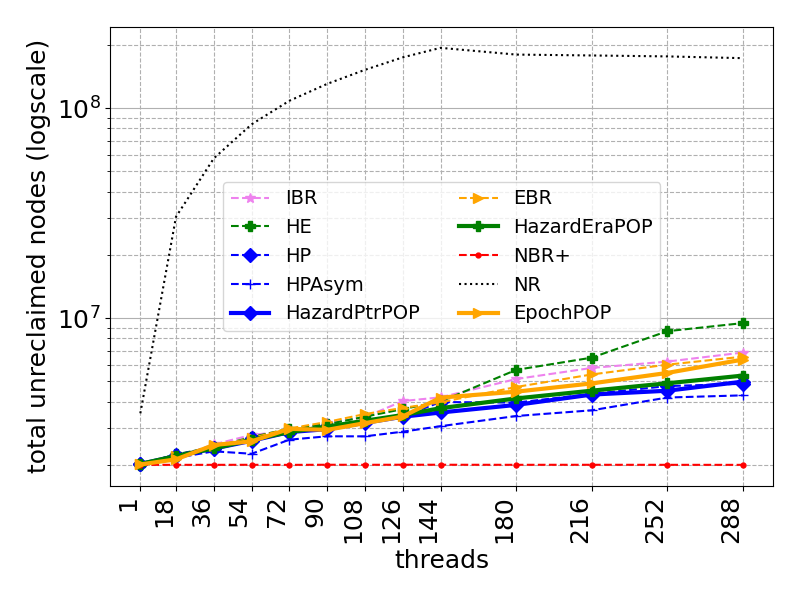}\hfill
            \caption{update heavy: 50\% inserts and 50\% deletes. }
            \label{fig:dgt100u}
        \end{subfigure}
        \begin{subfigure}{\textwidth}
            \includegraphics[width=0.33\linewidth, keepaspectratio]{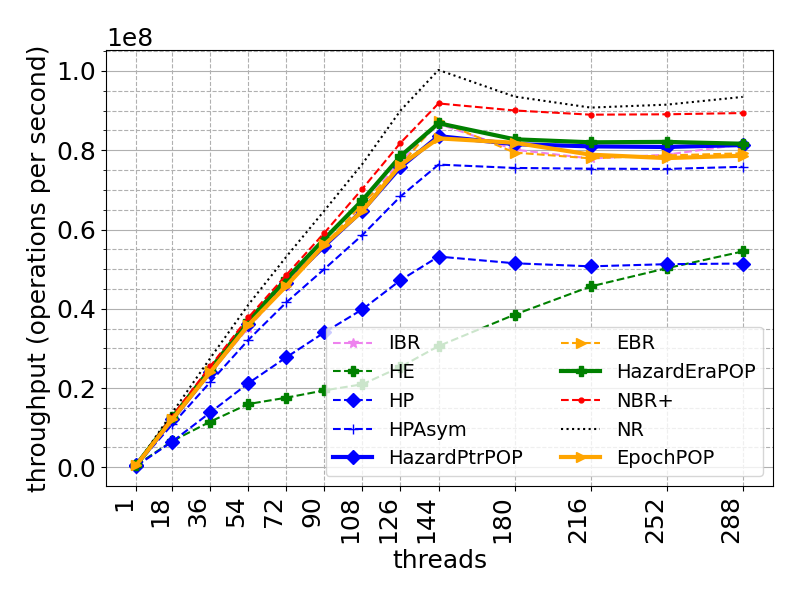}\hfill
            \includegraphics[width=0.33\linewidth, keepaspectratio]{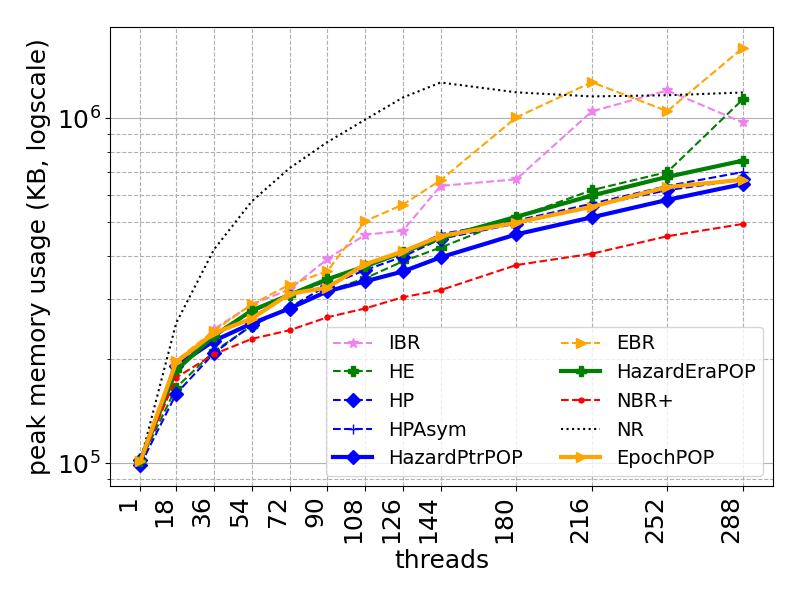}\hfill
            \includegraphics[width=0.33\linewidth, keepaspectratio]{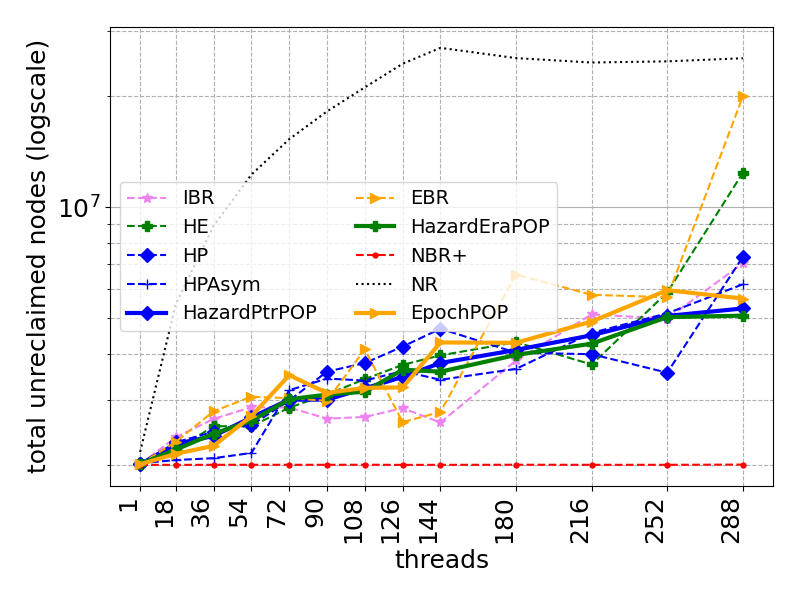}\hfill
            \caption{read heavy: 5\% inserts, 5\% deletes and 90\% contains.}
            \label{fig:dgt10u}
        \end{subfigure}
     \end{minipage}
    \label{fig:dgt}
    \caption{DGT. Size 2M. [Left: Throughput]. [Center: max resident memory in system]. [Right:total unreclaimed nodes].}
\end{figure*}

\begin{figure*}[h]
\centering
     \begin{minipage}{\textwidth}
        \begin{subfigure}{\textwidth}
            \includegraphics[width=0.33\linewidth, keepaspectratio]{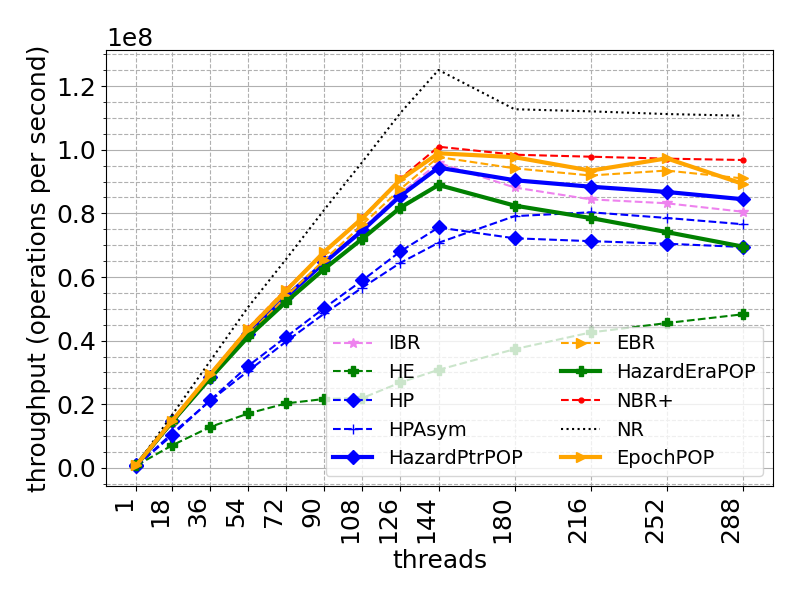}\hfill
            \includegraphics[width=0.33\linewidth, keepaspectratio]{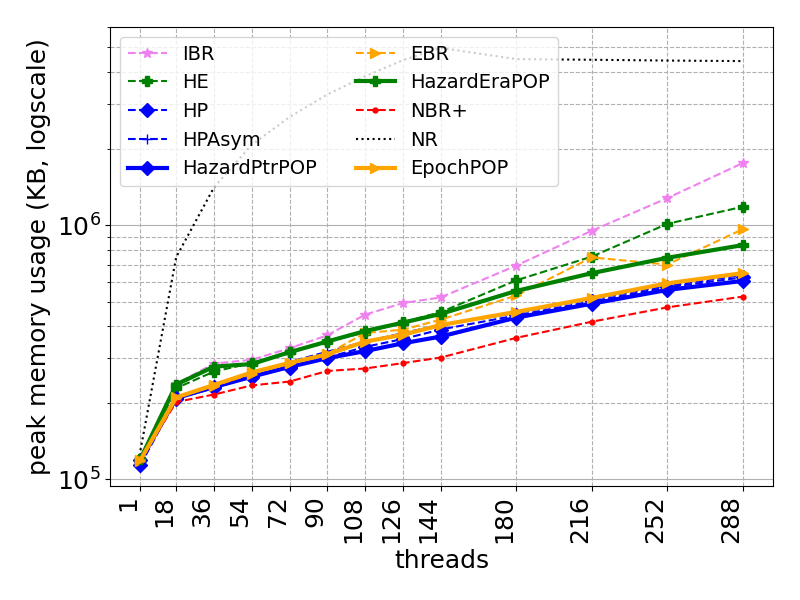}\hfill
            \includegraphics[width=0.33\linewidth, keepaspectratio]{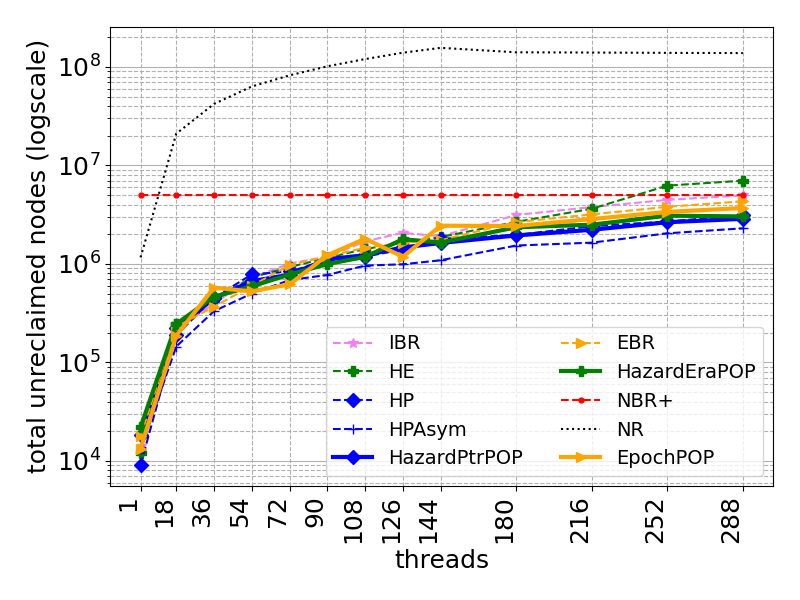}\hfill
            \caption{update heavy: 50\% inserts and 50\% deletes. }
            \label{fig:hmht100u}
        \end{subfigure}
        \begin{subfigure}{\textwidth}
            \includegraphics[width=0.33\linewidth, keepaspectratio]{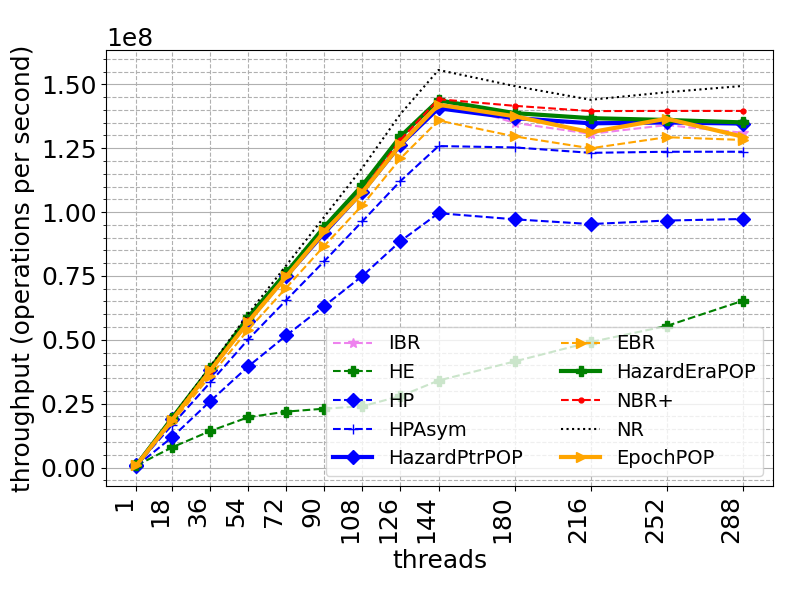}\hfill
            \includegraphics[width=0.33\linewidth, keepaspectratio]{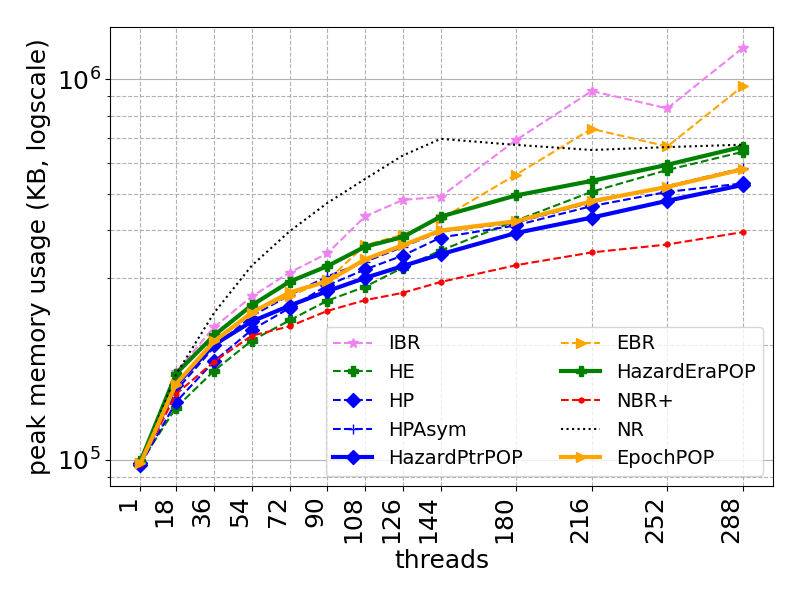}\hfill
            \includegraphics[width=0.33\linewidth, keepaspectratio]{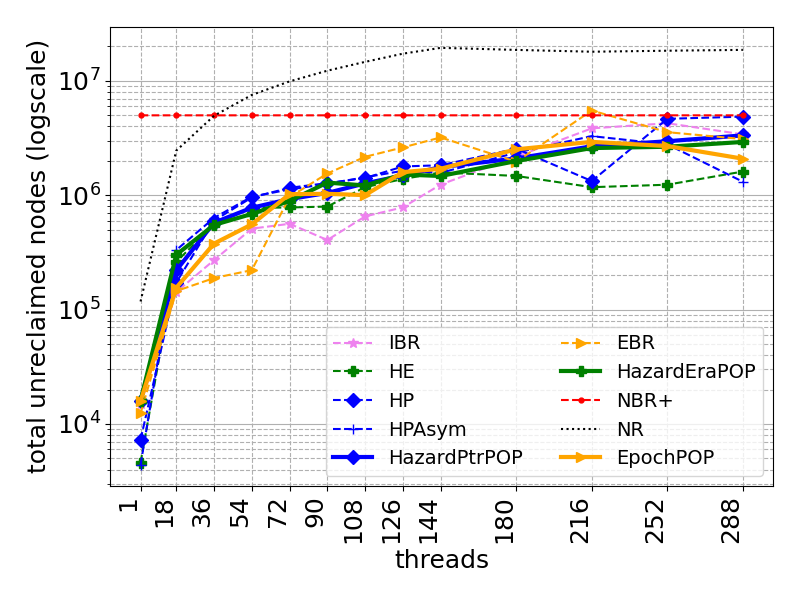}\hfill
            \caption{read heavy: 5\% inserts, 5\% deletes and 90\% contains.}
            \label{fig:hmht10u}
        \end{subfigure}
     \end{minipage}
    \label{fig:hmht}
    \caption{HT. Size 6M. [Left: Throughput]. [Center: max resident memory in system]. [Right:total unreclaimed nodes].}
\end{figure*}

\begin{figure*}[h]
\centering
     \begin{minipage}{\textwidth}
        \begin{subfigure}{\textwidth}
            \includegraphics[width=0.33\linewidth, keepaspectratio]{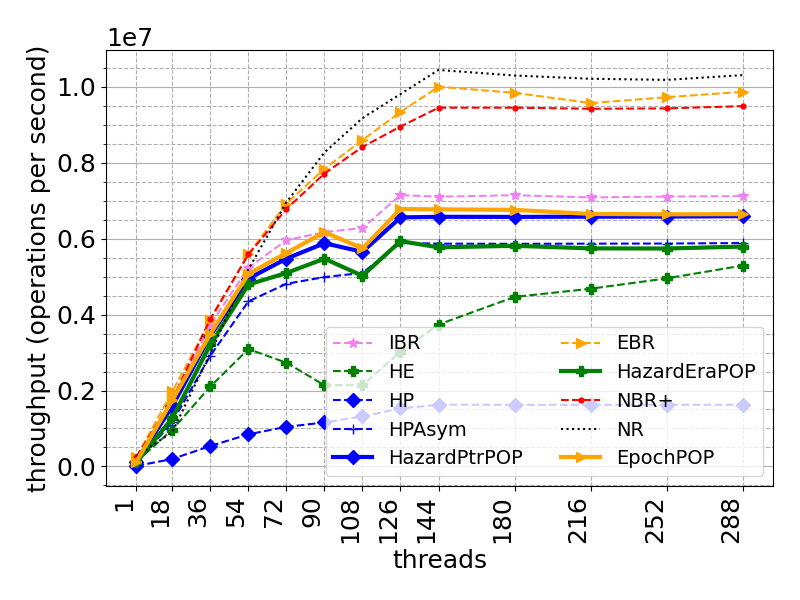}\hfill
            \includegraphics[width=0.33\linewidth, keepaspectratio]{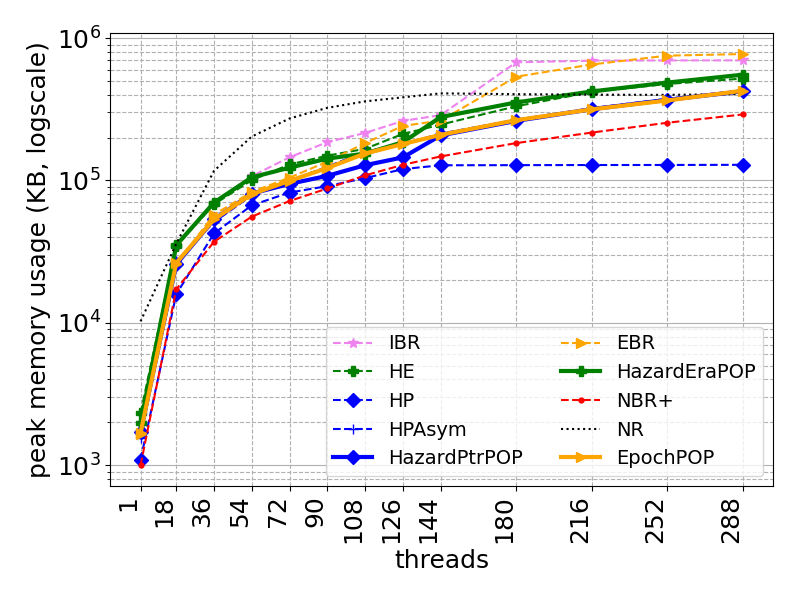}\hfill
            \includegraphics[width=0.33\linewidth, keepaspectratio]{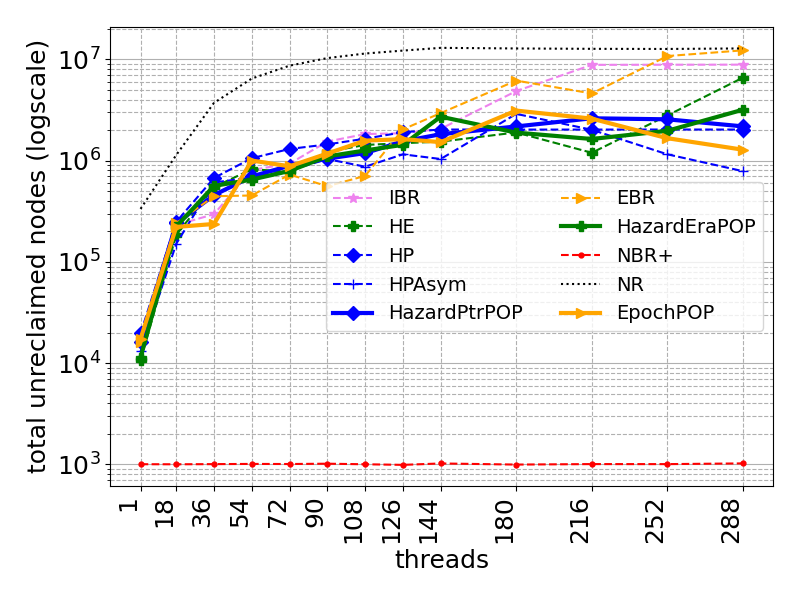}\hfill
            \caption{update heavy: 50\% inserts and 50\% deletes. }
            \label{fig:hml100u}
        \end{subfigure}
        \begin{subfigure}{\textwidth}
            \includegraphics[width=0.33\linewidth, keepaspectratio]{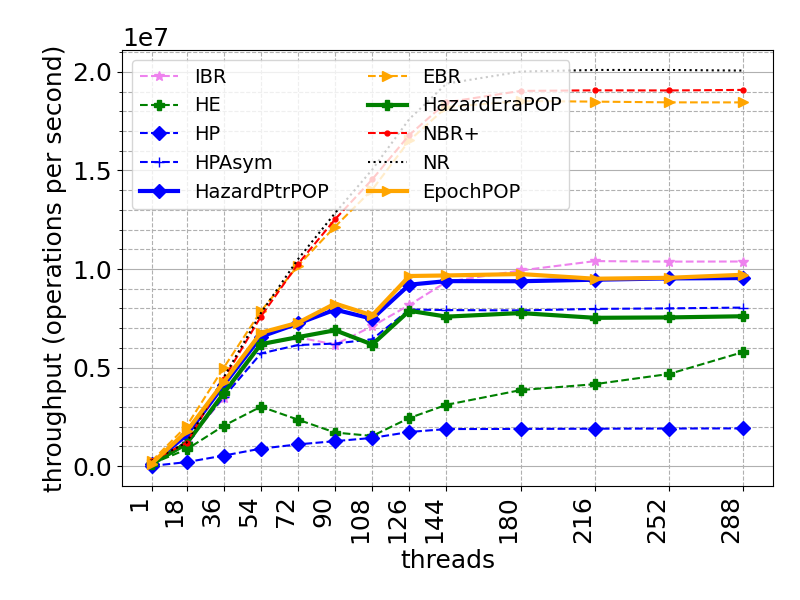}\hfill
            \includegraphics[width=0.33\linewidth, keepaspectratio]{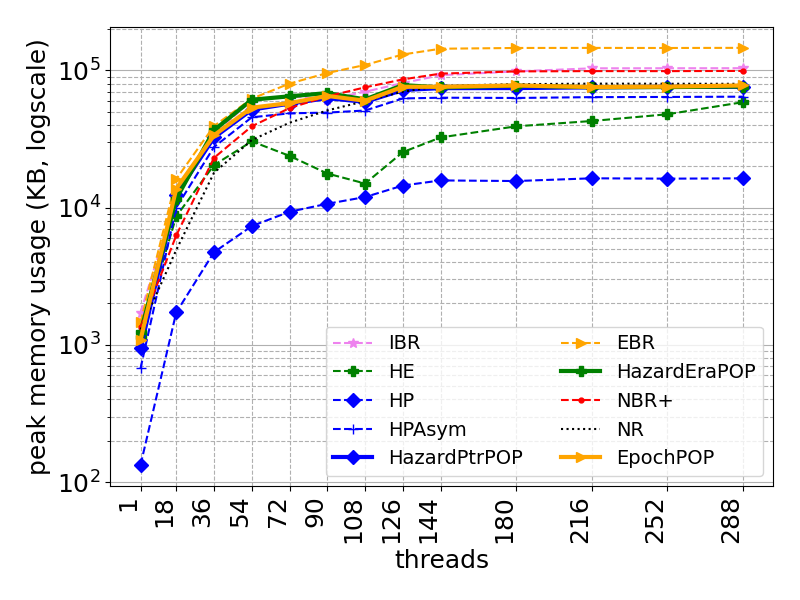}\hfill
            \includegraphics[width=0.33\linewidth, keepaspectratio]{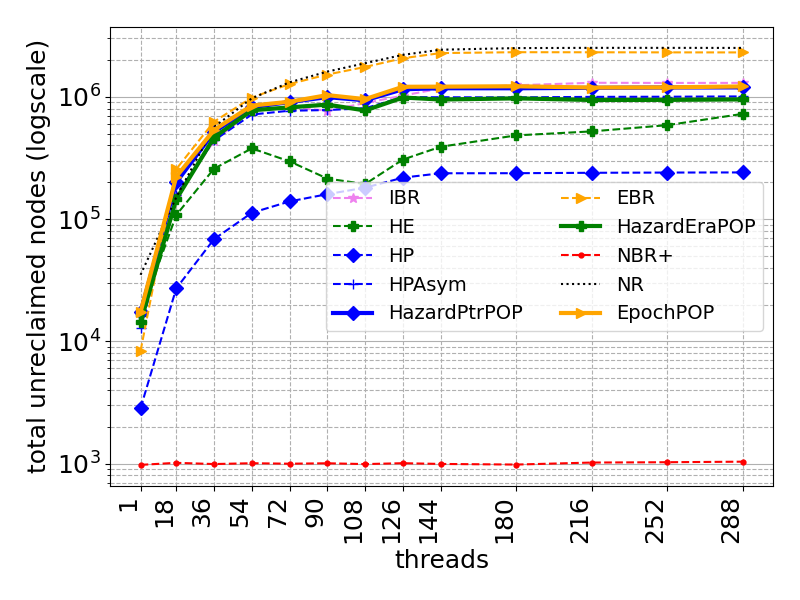}\hfill
            \caption{read heavy: 5\% inserts, 5\% deletes and 90\% contains.}
            \label{fig:hml10u}
        \end{subfigure}
     \end{minipage}
    \label{fig:hml}
    \caption{HML List. Size 2K. [Left: Throughput]. [Center: max resident memory in system]. [Right:total unreclaimed nodes].}
\end{figure*}

\begin{figure*}[h]
\centering
     \begin{minipage}{\textwidth}
        \begin{subfigure}{\textwidth}
            \includegraphics[width=0.33\linewidth, keepaspectratio]{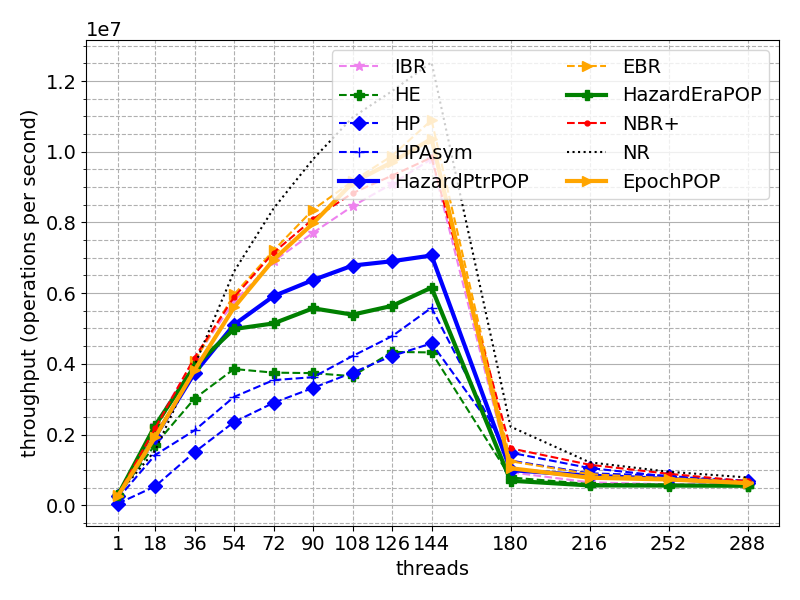}\hfill
            \includegraphics[width=0.33\linewidth, keepaspectratio]{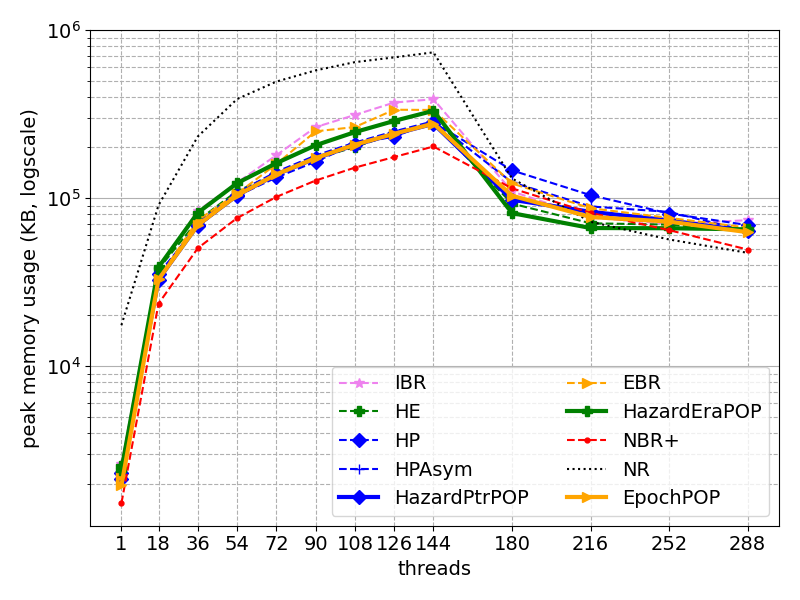}\hfill
            \includegraphics[width=0.33\linewidth, keepaspectratio]{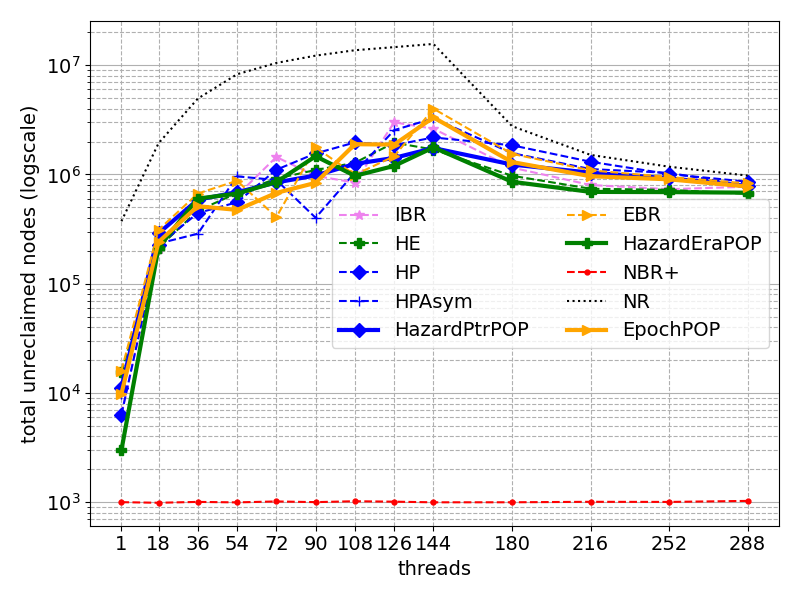}\hfill
            \caption{update heavy: 50\% inserts and 50\% deletes. }
            \label{fig:ll100u}
        \end{subfigure}
        \begin{subfigure}{\textwidth}
            \includegraphics[width=0.33\linewidth, keepaspectratio]{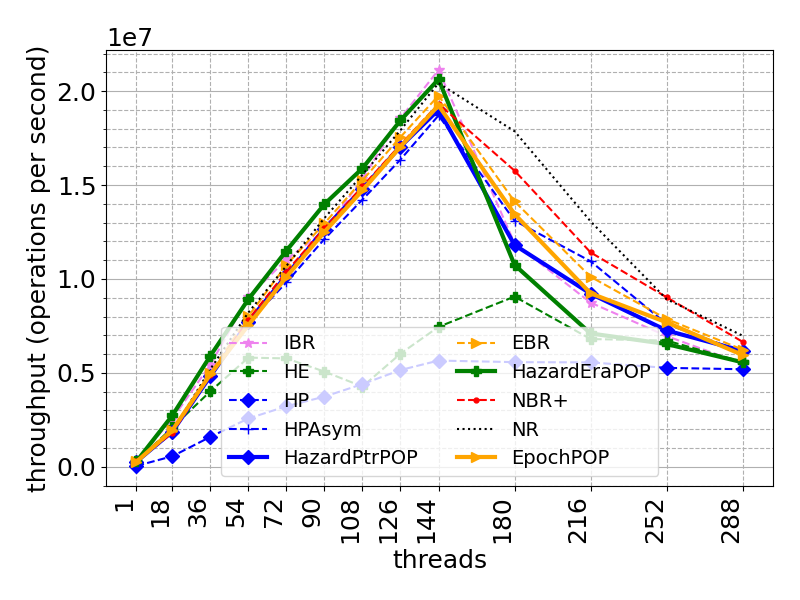}\hfill
            \includegraphics[width=0.33\linewidth, keepaspectratio]{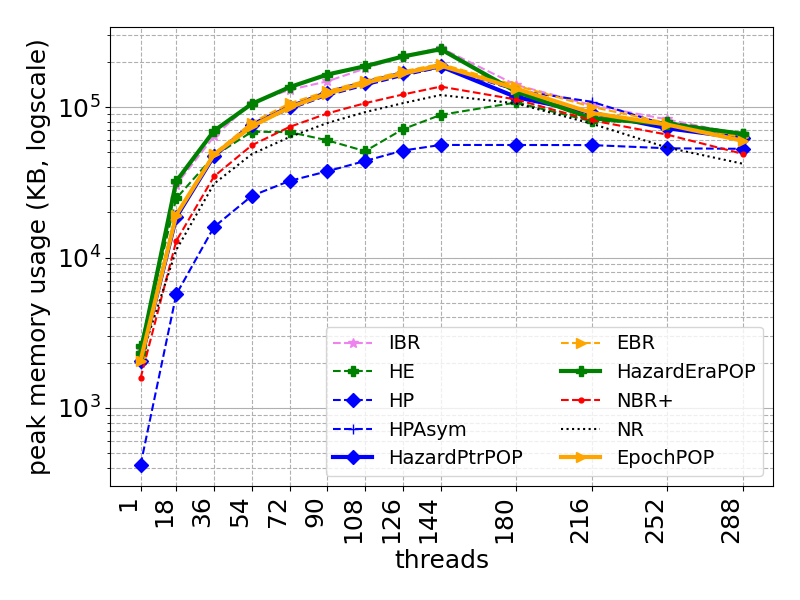}\hfill
            \includegraphics[width=0.33\linewidth, keepaspectratio]{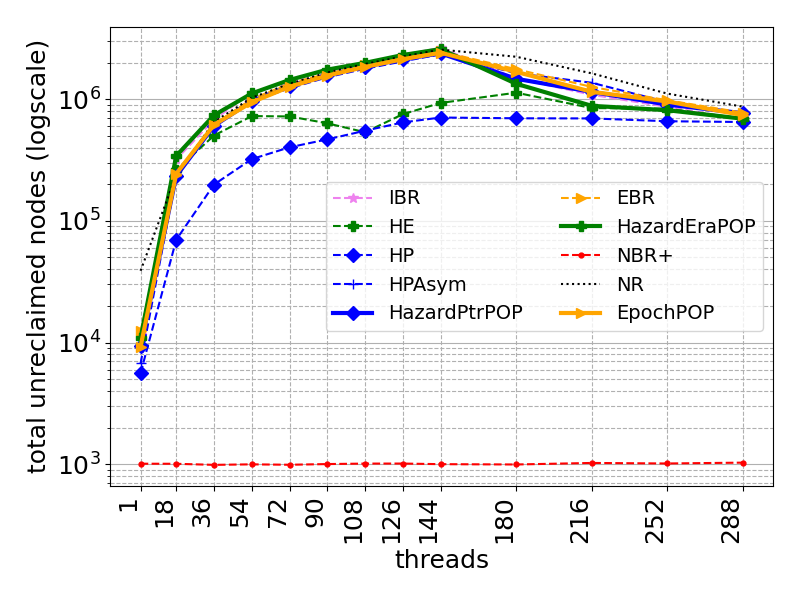}\hfill
            \caption{read heavy: 5\% inserts, 5\% deletes and 90\% contains.}
            \label{fig:ll10u}
        \end{subfigure}
     \end{minipage}
    \label{fig:ll}
    \caption{LL. Size 2K. [Left: Throughput]. [Center: max resident memory in system]. [Right:total unreclaimed nodes].}
\end{figure*}

\clearpage
\section{Experiments with Crystalline Algorithm~\cite{nikolaev2021crystalline}}
\label{sec:cyexp}

\begin{figure}[h]
\centering
     \begin{minipage}{\textwidth}
        \begin{subfigure}{\textwidth}
            \includegraphics[width=0.33\linewidth, keepaspectratio]{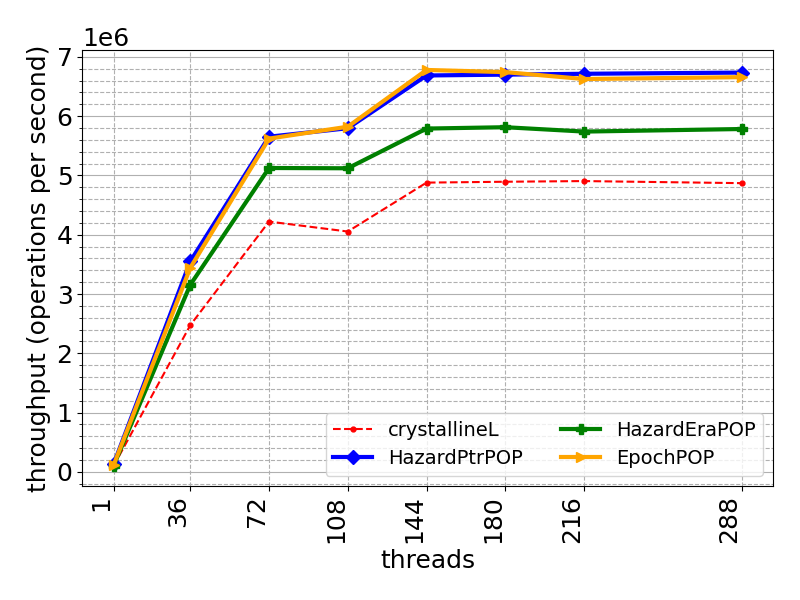}\hfill
            \includegraphics[width=0.33\linewidth, keepaspectratio]{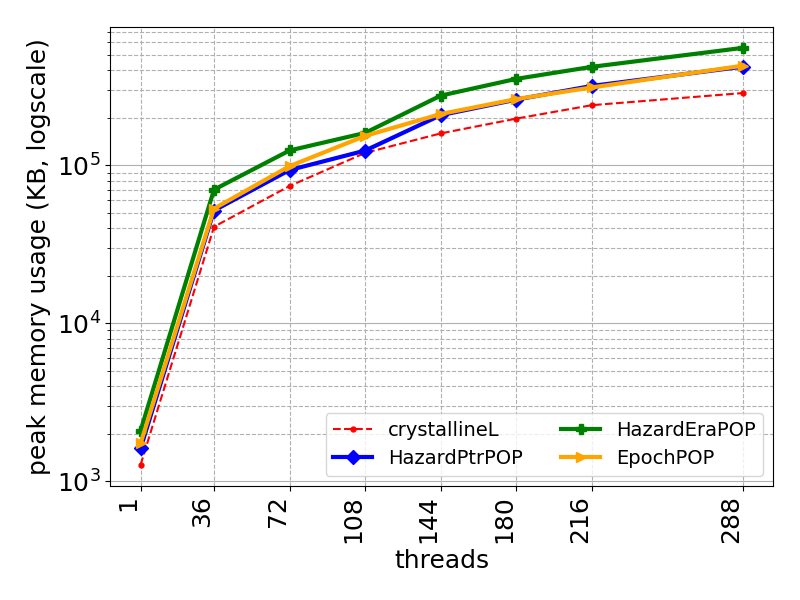}\hfill
            \includegraphics[width=0.33\linewidth, keepaspectratio]{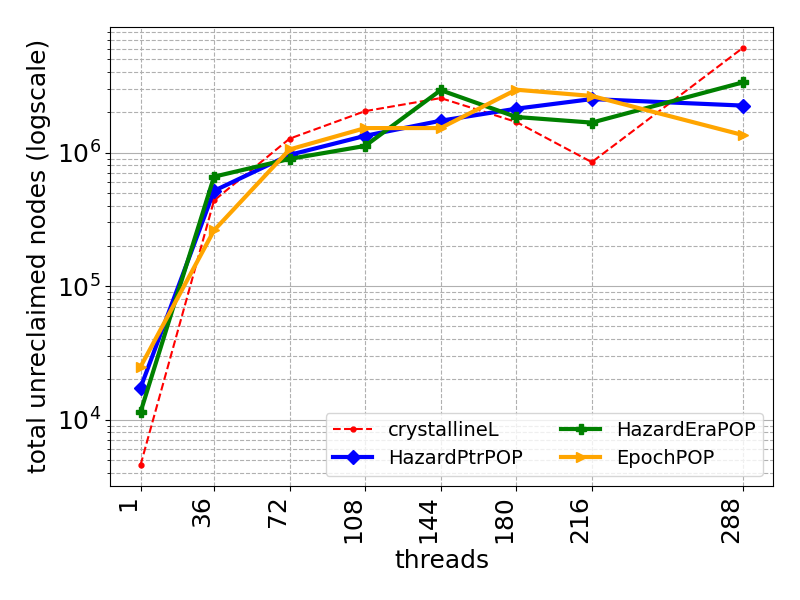}\hfill
            \caption{update heavy: 50\% inserts and 50\% deletes. }
            \label{fig:hml100u}
        \end{subfigure}
        \begin{subfigure}{\textwidth}
            \includegraphics[width=0.33\linewidth, keepaspectratio]{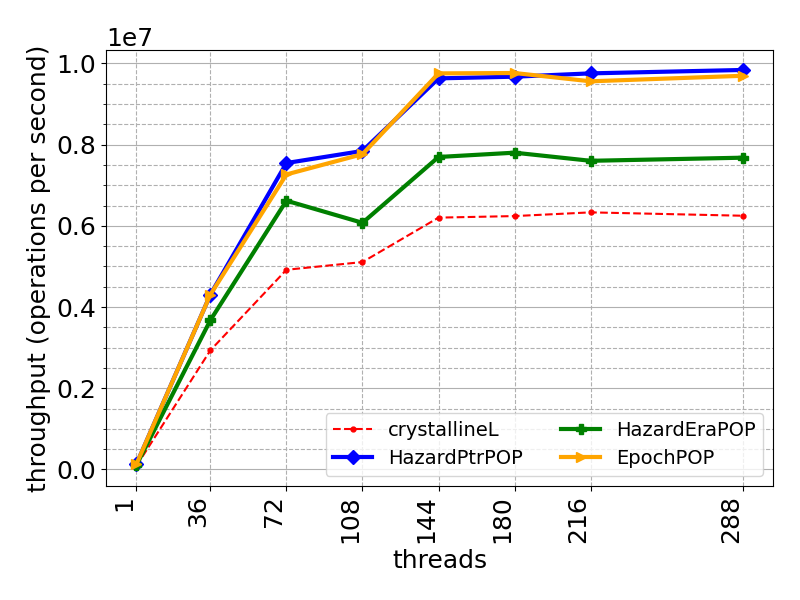}\hfill
            \includegraphics[width=0.33\linewidth, keepaspectratio]{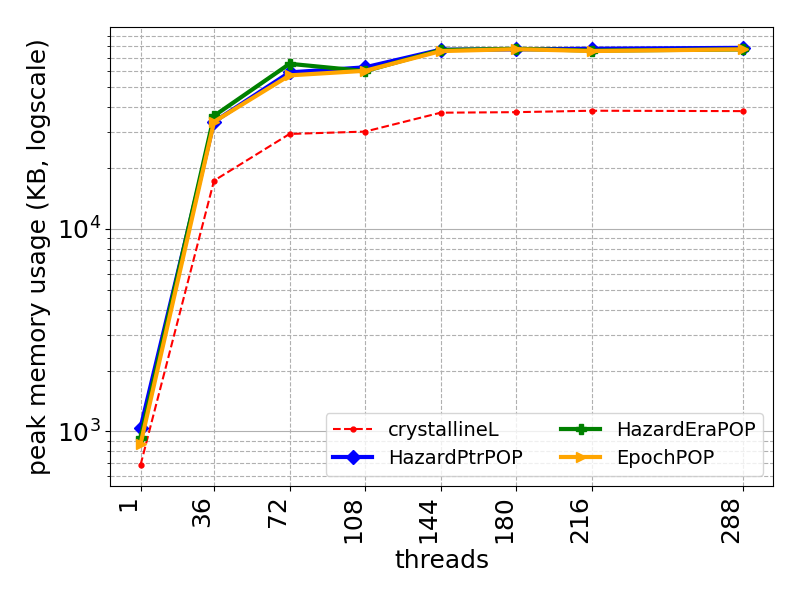}\hfill
            \includegraphics[width=0.33\linewidth, keepaspectratio]{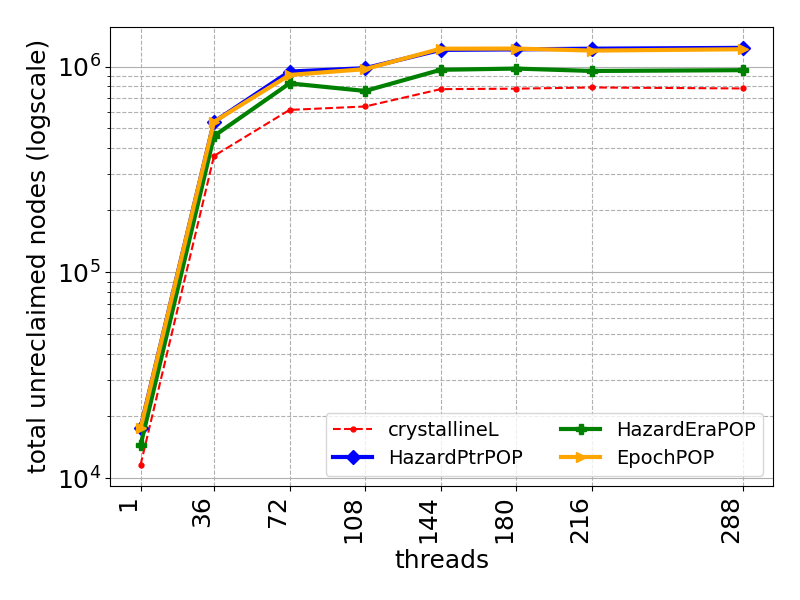}\hfill
            \caption{read heavy: 5\% inserts, 5\% deletes and 90\% contains.}
            \label{fig:hml10u}
        \end{subfigure}
     \end{minipage}
    \label{fig:hml}
    \caption{HML List. Size 2K. [Left: Throughput]. [Center: max resident memory in system]. [Right:total unreclaimed nodes].}
\end{figure}
\begin{figure}[h]
\centering
     \begin{minipage}{\textwidth}
        \begin{subfigure}{\textwidth}
            \includegraphics[width=0.33\linewidth, keepaspectratio]{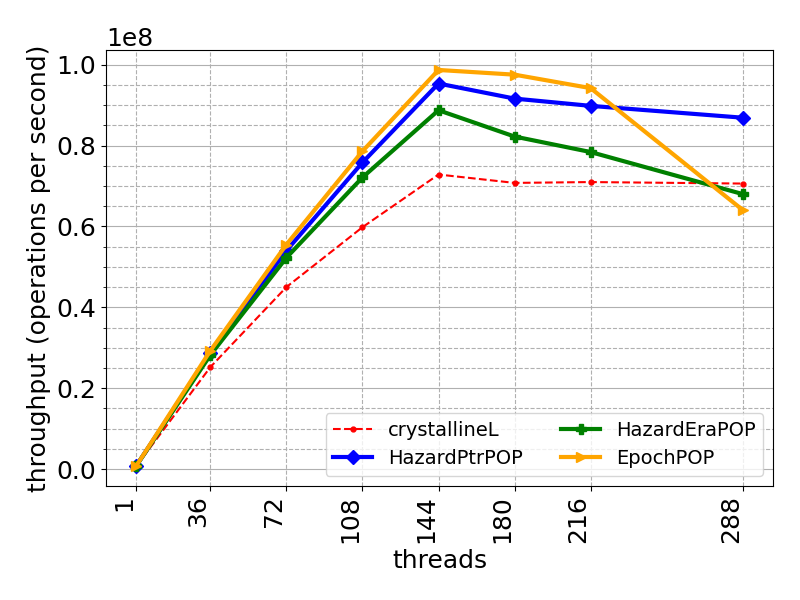}\hfill
            \includegraphics[width=0.33\linewidth, keepaspectratio]{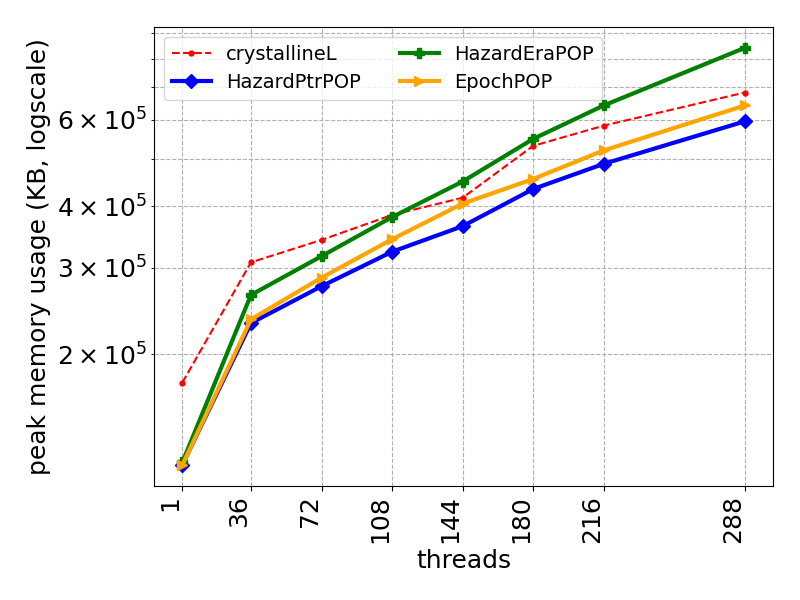}\hfill
            \includegraphics[width=0.33\linewidth, keepaspectratio]{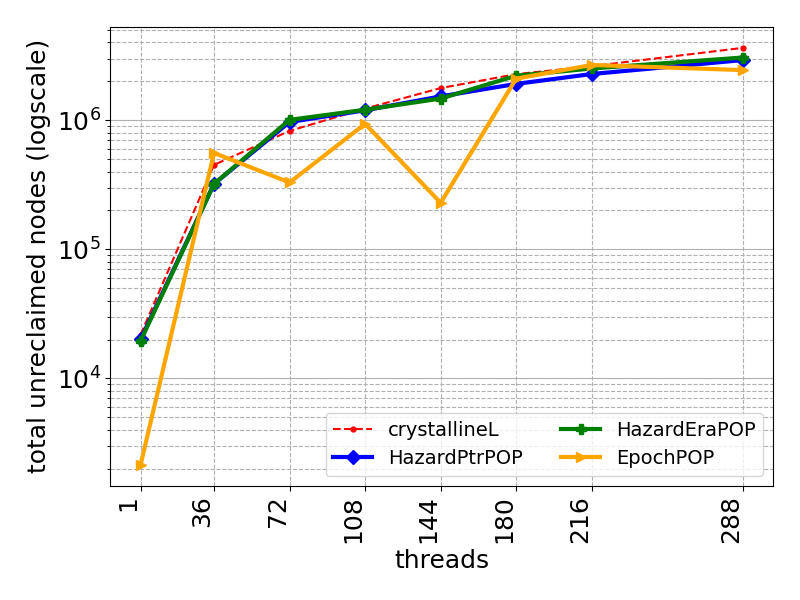}\hfill
            \caption{update heavy: 50\% inserts and 50\% deletes. }
            \label{fig:hmht100u}
        \end{subfigure}
        \begin{subfigure}{\textwidth}
            \includegraphics[width=0.33\linewidth, keepaspectratio]{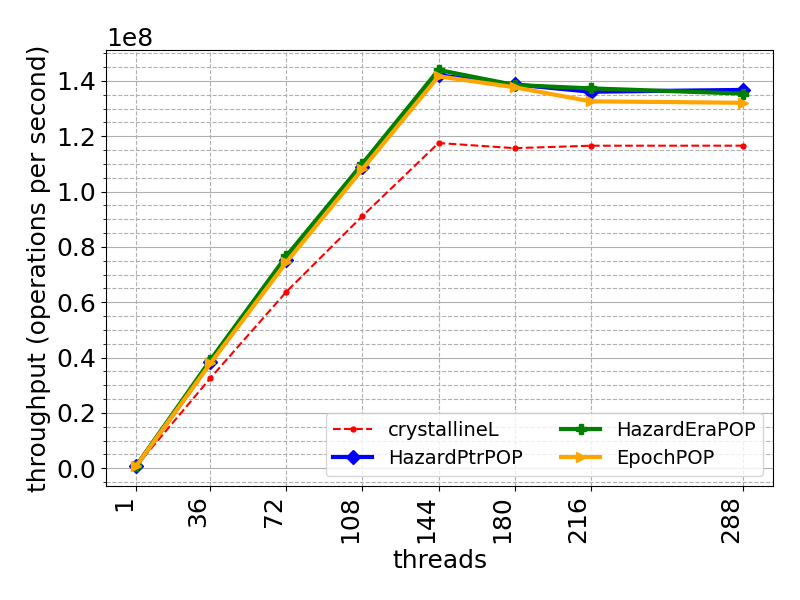}\hfill
            \includegraphics[width=0.33\linewidth, keepaspectratio]{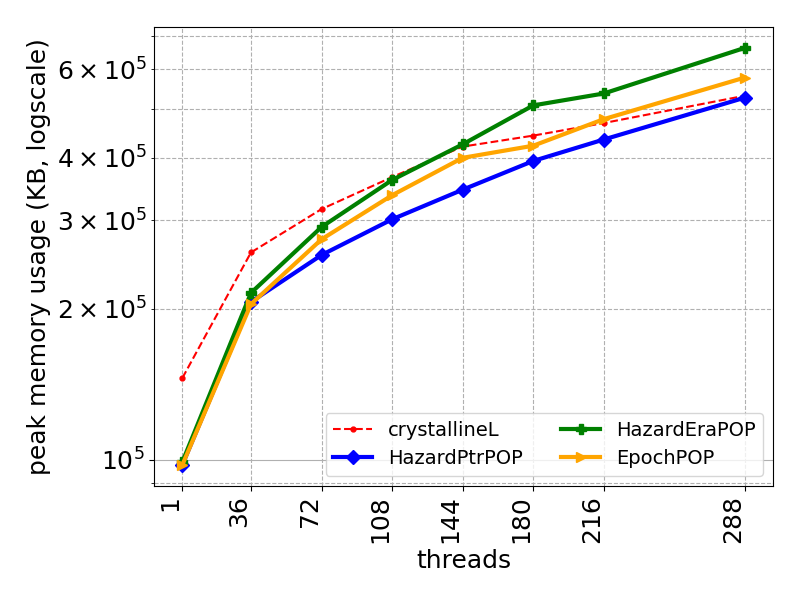}\hfill
            \includegraphics[width=0.33\linewidth, keepaspectratio]{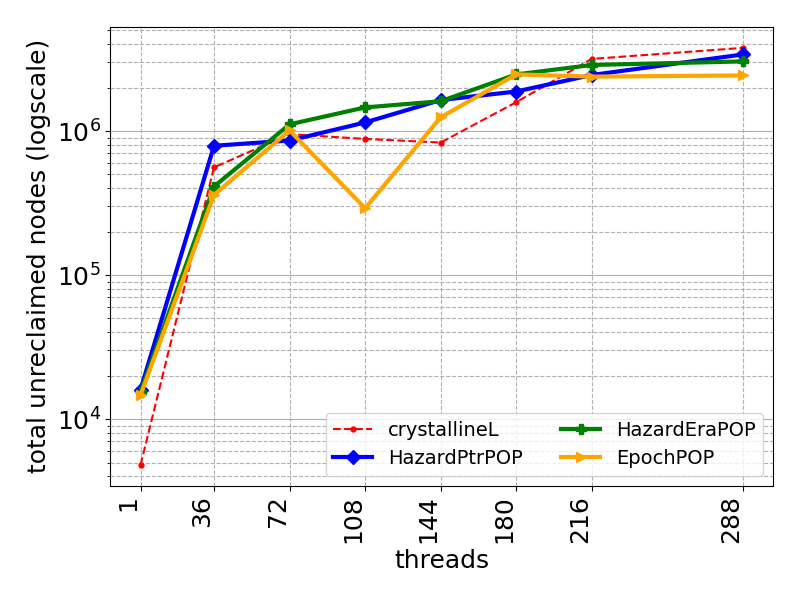}\hfill
            \caption{read heavy: 5\% inserts, 5\% deletes and 90\% contains.}
            \label{fig:hmht10u}
        \end{subfigure}
     \end{minipage}
    \label{fig:hmht}
    \caption{HT. Size 6M. [Left: Throughput]. [Center: max resident memory in system]. [Right:total unreclaimed nodes].}
\end{figure}
\end{document}